\documentclass[11pt,a4paper]{article}
\usepackage[utf8]{inputenc}
\usepackage{graphicx}
\usepackage{float}
\usepackage{afterpage}
\usepackage{epstopdf}
\usepackage{epsfig,cite}
\usepackage{amssymb}
\usepackage{amsmath}
\usepackage{dsfont}
\usepackage{multirow}
\usepackage{url}
\usepackage[colorlinks=false]{hyperref}
\usepackage{subfig}
\usepackage{listings}
\usepackage{amsmath}

\usepackage{url}
\usepackage{hyperref}

\usepackage [autostyle, english = american]{csquotes}
\MakeOuterQuote{"}
\newcommand{\be}{\begin{equation}}
\newcommand{\ee}{\end{equation}}
\newcommand{\bea}{\begin{eqnarray}}
\newcommand{\eea}{\end{eqnarray}}
\newcommand{\bi}{\begin{itemize}}
\newcommand{\ei}{\end{itemize}}
\newcommand{\ben}{\begin{enumerate}}
\newcommand{\een}{\end{enumerate}}
\newcommand{\la}{\left\langle}
\newcommand{\ra}{\right\rangle}
\newcommand{\lc}{\left[}
\newcommand{\rc}{\right]}
\newcommand{\lp}{\left(}
\newcommand{\rp}{\right)}

\def\frac#1#2{{{#1}\over {#2}}}
\def\gsim{\mathrel{\rlap{\lower4pt\hbox{\hskip1pt$\sim$}}
    \raise1pt\hbox{$>$}}}         
\def\lsim{\mathrel{\rlap{\lower4pt\hbox{\hskip1pt$\sim$}}
    \raise1pt\hbox{$<$}}}         

\newcommand{\draft}[1]{}

\def\beq{\begin{equation}}  
\def\eeq{\end{equation}}  

\def \n0{N_j^{(0)}}

\def\lapprox{\lower .7ex\hbox{$\;\stackrel{\textstyle <}{\sim}\;$}}
\def\gapprox{\lower .7ex\hbox{$\;\stackrel{\textstyle >}{\sim}\;$}}

\begin{document}

\begin{figure}[h]
\epsfig{width=0.35\textwidth,figure=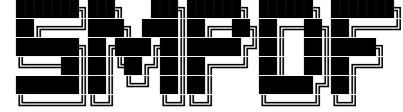}
\end{figure}

\begin{flushright}
  CERN-PH-TH-2015-243\\
  TIF-UNIMI-2015-13\\
  OUTP-15-24P \\ 
  \end{flushright}

\vspace{0.4cm}

\begin{center}
  {\Large \bf Specialized minimal PDFs for optimized LHC calculations}
\vspace{.7cm}

Stefano Carrazza$^{1,2}$, Stefano Forte$^1$, Zahari Kassabov$^{1,3}$ and
Juan Rojo$^4$

\vspace{.3cm}
{\it ~$^1$ TIF Lab, Dipartimento di Fisica, Universit\`a di Milano and
  INFN, Sezione di Milano,\\ Via Celoria 16, I-20133 Milano, Italy\\
  ~$^2$ Theory Department, CERN, CH-1211 Geneva 23, Switzerland \\
~$^3$ Dipartimento di Fisica, Universit\`a di Torino and
INFN, Sezione di Torino,\\ Via Pietro Giuria 1, I-10125 Torino, Italy\\
~$^4$ Rudolf Peierls Centre for Theoretical Physics, 1 Keble Road,\\ University of Oxford, OX1 3NP Oxford, United Kingdom\\}
\end{center}   

\vspace{0.1cm}

\begin{center}
{\bf \large Abstract}

\end{center} 
We present a methodology for the construction 
of parton distribution functions (PDFs) designed to provide an accurate
representation of PDF  uncertainties for specific processes or classes
of processes with a minimal number of PDF
error sets: specialized minimal PDF sets, or SM-PDFs.
We construct these  SM-PDFs
in such a way that
 sets corresponding to different
 input processes can be combined without losing information,
 specifically on their 
 correlations, and that they are
 robust upon smooth variations of
 the 
 kinematic cuts.
 The proposed strategy never discards information, so that
 the SM-PDF sets can be enlarged by the addition of new
 processes, until the prior PDF set is eventually
 recovered for a large enough set of processes.
We illustrate
the method by producing SM-PDFs tailored to Higgs, top quark
pair, and electroweak gauge boson physics,
and determine that,
when the PDF4LHC15 combined
set is used as the prior,
around 11, 4 and 11 Hessian
eigenvectors respectively are enough
to fully describe the corresponding processes.
\clearpage

\tableofcontents

\section{Introduction}
\label{sec:Introduction}

Modern sets of parton distributions (PDFs)~\cite{Alekhin:2013nda,Dulat:2015mca,Owens:2012bv,Abramowicz:2015mha,Harland-Lang:2014zoa,Ball:2014uwa}
provide a representation of their associated
uncertainties based on either the Hessian~\cite{Pumplin:2001ct} or
the Monte Carlo (MC)~\cite{DelDebbio:2004qj}
methods, supplementing their central PDF member with additional
error members (eigenvectors
or MC replicas).
The number of PDF members required for an accurate
representation of PDF uncertainty can be as large as several hundreds,
especially when constructing PDF sets based on the combination of
several underlying PDFs fitted to data: for example, the recent
PDF4LHC 2015  sets~\cite{Butterworth:2015oua} are based
on a combined sample of 900 MC PDF replicas.

The usage of such large PDF samples can be computationally unwieldy, 
and this motivated the development of  strategies for reducing the 
number of  PDF members while minimizing accuracy loss.
A number of such reduction strategies have been made
available recently.
Two of these methods provide a  Hessian representation
of the prior PDF set
in terms of a smaller number of eigenvectors:  META-PDFs~\cite{Gao:2013bia},
and MCH-PDFs~\cite{Carrazza:2015aoa}.
A third method uses a compression algorithm to reduce
the number of replicas of a underlying MC PDF prior: CMC-PDFs~\cite{Carrazza:2015hva}.

These three methods have been extensively benchmarked in the context
of the 2015 PDF4LHC recommendations~\cite{Butterworth:2015oua}, where
it was found 
that generally a set of about a hundred PDFs is required in order to
represent PDF uncertainties with percentage accuracy
for all PDFs in the complete range of $(x,Q)$ relevant
for LHC phenomenology.
However, it is well known~\cite{Pumplin:2009bb} that, if one is
interested only in the description of a specific
set of cross-sections, the number
of PDF error members  can be
greatly reduced without significant accuracy loss.

In this work
we propose a new strategy to achieve this goal. Our 
methodology, which we denote by Specialized Minimal PDFs (SM-PDFs),
is based on the Singular Value Decomposition 
version of the  {\tt mc2hessian} algorithm,
as presented in the Appendix of Ref.~\cite{Carrazza:2015aoa}.
Starting from a either a Hessian or
a Monte Carlo prior set
and a list of collider processes,
the SM-PDF algorithm leads to a set of eigenvectors
optimized  for the description of the input processes
within some given tolerance. 

In comparison to existing methods, such as
data set diagonalization~\cite{Pumplin:2009bb}, our methodology has
the advantage that no information is lost in the process of the
construction of the specialized set.
This is because the specialized set is
constructed through a suitable linear transformation, whereby the
starting space is separated into a subspace spanned by the optimized
SM-PDF set, and its orthogonal subspace.
This then implies that
any given SM-PDF set can be iteratively expanded in order to maintain
a given accuracy for an increasingly large set of processes, and also, 
that SM-PDF sets optimized for different sets of processes can be
combined into a single set, either {\it a priori}, at the level
of PDFs, or {\it a posteriori},
at the level of cross-sections.
This, for example, enables the a-posteriori combination of previous
 independent studies for a signal process and its corresponding 
backgrounds, with all
correlations properly accounted for.

This paper is organized as follows: In Sect.~\ref{sec:methodology} we
describe our general
strategy and methodology in detail.
Then, in
Sect.~\ref{sec:validation} we apply our method to the most important
Higgs production channels ($ggh$, $ht\bar{t}$ and $hV$, VBF $h$) as well as
for other standard candles at the LHC, \textit{i.e.} $t\bar{t}$, $Z$ and $W$ production.
We
compute one specific reduced sets for each of them, as well as as
single set for all the processes combined.
We validate the
results by comparing the predictions of these reduced sets to the
prior input set. We also show that our method provides an adequate
generalization by showing that the predictions are stable when
computing similar processes but with different kinematical cuts than
those used as input.
In Sect.~\ref{sec:combination} we show how
experimental analyses done with different SM-PDFs can be 
combined together.
In Sect.~\ref{sec:delivery} we provide an overview
of the deliverables of this work, in particular the code itself which
allows to easily generate reduced sets with personalized configuration
and the {\tt LHAPDF6}~\cite{Buckley:2014ana} sets of SM-PDFs for the
processes described in Sect.~\ref{sec:delivery}.
Finally, Appendix~\ref{sec-appendix-correlations} presents a
graphical illustration of the regions in PDF space which give the
dominant contribution to various physical processes,
 and Appendix~\ref{sec-appendix}
 provides some basic instructions for the execution of the
 SM-PDF code.

\section{Methodology}
\label{sec:methodology}
The  SM-PDF methodology is built upon the strategy based on Singular-Value
Decomposition (SVD) followed by
Principal Component Analysis (PCA)
described in the Appendix of Ref.~\cite{Carrazza:2015aoa}, in which the
MCH method was presented.
This SVD+PCA 
strategy achieves the twofold goal of obtaining a
multigaussian representation of a starting (prior) Monte Carlo PDF
set, and of allowing for an optimization of this representation
for a specific set of input cross-sections, which
uses the minimal number of eigenvectors required in order
to reach a desired accuracy goal.
We will now review the SVD+PCA method, and
describe how it can be used for  the construction of specialized
minimal PDF
sets, optimized for the description
of a specific set of cross sections.

\subsection{The SVD+PCA method}
\label{sec:svd-pca}

The main problem we are addressing is the faithful representation of
PDF uncertainties, which typically requires a large number of PDF
error or Monte Carlo sets.
Here we will assume the 
central value to be the same as in the prior PDF set, from which, if
the prior is given as a Monte Carlo, it is typically determined as a
mean (though different choices, such as the median, are possible and
might be advisable in particular circumstances).

Hence, we are interested in the construction of a 
multigaussian representation in PDF space: the only information we need is
then the corresponding covariance matrix.
This is constructed starting
with  a  matrix $X$ which samples over a grid of points the difference between
each PDF replica,
$f_{\alpha}^{(k)}(x_{i},Q)$,
and the central set,
$f_{\alpha}^{(0)}(x_{i},Q)$, namely
\begin{equation}
  X_{lk}(Q)\equiv
  f_{\alpha}^{(k)}(x_{i},Q)-f_{\alpha}^{(0)}(x_{i},Q)\,, \label{eq:Xmat}
\end{equation}
where $\alpha$ runs over the $N_f$ independent PDF flavors at the
factorization scale $\mu_F=Q$, $i$ runs over the $N_x$ points in the
$x$ grid where the PDFs are sampled, $l= N_{x}(\alpha-1)+i$ runs
over all $N_{x}N_{f}$ grid points, and $k$ runs over the
$N_{\textrm{rep}}$ replicas. The sampling is chosen to be fine-grained
enough that  results will not depend on it.

The desired covariance matrix in PDF space is then constructed as
\begin{equation}
  \label{covmat}
\textrm{cov}(Q) =
\frac{1}{N_{\rm rep}-1}XX^t \ .
\end{equation}
The key idea which underlies the SVD method is to represent the
$(N_{x}N_{f})\times(N_{x}N_{f})$ covariance matrix
Eq.~(\ref{covmat})
over the $N_{\rm rep}$ dimensional
linear space spanned by the replicas (assuming $N_{\rm
  rep}>N_{x}N_{f}$), by viewing its $N_{x}N_{f}$ eigenvectors as
orthonormal basis vectors in this
space, which can thus be represented as linear combinations of
replicas.
The subsequent PCA optimization then simply consists of
picking the subspace spanned by the dominant eigenvectors, {\it i.e.}, those
with largest eigenvalues.

The first step is the SVD of
the sampling matrix $X$, namely
\begin{equation}
  \label{eq:svd}
X=USV^t\ ,  
\end{equation}
where $U$ and $V^t$ are orthogonal matrices, with dimensions respectively
$N_x N_f\times N_{\rm eig}^{(0)}$ and  $N_{\rm rep} \times N_{\rm rep}$,
$S$ is a diagonal $N^{(0)}_{\rm eig}\times
N_{\rm rep}$ 
positive semi-definite matrix, whose elements are the so-called singular values of
$X$, and the initial number of singular values is given by  $N^{(0)}_{\rm  eig}=N_xN_f$.
Note that, because $S$ is diagonal, it can be
equivalently viewed as a $N^{(0)}_{\rm eig}\times
N^{(0)}_{\rm eig}$ matrix, since 
(with $N^{(0)}_{\rm eig}>N_{\rm  rep}$) all its further entries vanish.
This point of view was taken in the Appendix of~\cite{Carrazza:2015aoa}.
In this  case, only the 
$N^{(0)}_{\rm eig} \times N_{\rm rep}$ submatrix which actually
contributes to the SVD of the matrix $V$ is included.
 However, for the procedure to be described below, it is
more convenient to view $V$ as  $N_{\rm rep} \times N_{\rm rep}$
orthogonal matrix. 

  The matrix $Z=US$ then has the important property that 
\begin{equation}\label{svd1}
ZZ^t=X X^t,
\end{equation}
but also that it can be expressed as
\begin{equation}\label{svd2}
Z=XV,
\end{equation}
 and thus it provides the sought-for representation of the multigaussian
covariance matrix in terms of the original PDF replicas: specifically, $V_{kj}$ is the
expansion coefficient of  the $j$-th eigenvector over the $k$-th
replica.
We assume henceforth that the singular values are ordered, so
that the first diagonal entry of $S$ correspond to the largest value,
the second to the second-largest and so forth.

The PCA optimization then consists of only retaining the principal
components, {\it i.e.}  the largest singular values. In this case, 
$U$ and $S$  are replaced by their 
sub-matrices, denoted by $u$ and $s$ respectively,
with dimension $N_x N_f\times N_{\rm
  eig}$ and
$N_{\rm eig}\times
N_{\rm rep}$, with $N_{\rm eig} < N_{\rm  eig}^{(0)}$ the number of
eigenvectors which have been retained.
Due to the ordering,
these are the upper left sub-matrices. Because $s$ has only
$N_{\rm eig}$ non-vanishing diagonal entries, only the
$N_{\rm  rep}\times N_{\rm  eig}$
submatrix of $V$ contributes. We call  this the
principal  submatrix $P$ of  $V$:
\begin{equation}\label{eq:pdef}
  P_{kj}= V_{kj}\qquad  k=1\,,
  \dots,N_{\rm rep} \,,\qquad  j=1,\dots,N_{\rm eig}\, .
\end{equation}

The optimized representation of the original covariance matrix,
Eq.~(\ref{covmat}), is then found by replacing $V$
with its principal submatrix $P$ in Eq.~(\ref{svd2}).
This principal matrix $P$ is thus
the output of the SVD+PCA method: it contains
the coefficients of the linear combination of the original replicas or
error sets which correspond to the principal components, which can be
used to compute PDF uncertainties using the Hessian
method.

Indeed, given a certain observable $\sigma_i$ (which
could be a cross-section, the value of a
structure function, a bin of a differential distribution, etc.)
its PDF uncertainty can be computed
in terms of the original Monte Carlo replicas by
\begin{equation}
   \label{eq:obsstd}
   s_{\sigma_i} = \left(\frac{1}{N_{\rm rep}-1}\sum_{k=1}^{N_{\rm rep}}\left(\sigma_i^{(k)} -
   \sigma_i^{(0)}\right)^2\right)^\frac{1}{2}= \frac{1}{\sqrt{N_{\rm rep}-1}}\left\Vert d(\sigma_i)\right\Vert  ,
\end{equation}
where  $\sigma_i^{(k)}$ is the prediction obtained using the $k$-th
Monte Carlo PDF replica, $\sigma_i^{(0)}$ is the central prediction, and
in the last step we have defined the vector of
differences 
\begin{equation}
   \label{eq:obsdiffs}
   d_k(\sigma_i) \equiv  \sigma_i^{(k)} - \sigma_i^{(0)} \, ,\qquad
   k=1,\ldots,N_{\rm rep} \, ,
\end{equation}
with norm
\begin{equation}
   \label{eq:errvec}
 \left\Vert d(\sigma_i)\right\Vert \equiv \left(\sum_{k=1}^{N_{\rm
   rep}}d_k^2(\sigma_i) \right)^\frac{1}{2}
  .
\end{equation}

Assuming linear error propagation and using Eq.~(\ref{svd2}), 
the norm of the
vector  $\{ d_k(\sigma_i)\}$, Eq.~(\ref{eq:obsdiffs}), can be
represented on the eigenvector basis:
\begin{equation}\label{rotation}
\left\Vert d(\sigma_1)\right\Vert= 
\left\Vert {d^V}(\sigma_1)\right\Vert\end{equation}
where the rotated vector 
\begin{equation}
   \label{eq:diffpredicted}
   {d^V}_j(\sigma_i) = \sum_{k=1}^{N_{\rm rep}}
   d_k(\sigma_i)
   V_{kj}, \qquad j=1,\dots,N^{(0)}_{\rm eig} \, ,
\end{equation}
has the same norm as the original one because of Eq.~(\ref{svd1}).

Replacing $V$ by the principal matrix $P$ in Eq.~(\ref{eq:diffpredicted}),
{\it i.e.}, letting $j$ only run up to $N_{\rm eig}<N^{(0)}_{\rm eig}$
we get
\begin{equation}
   \label{eq:errfinal}
   \tilde{s}_{\sigma_i} = \frac{1}{\sqrt{N_{\rm rep}-1}}
\left\Vert {d^P}(\sigma_i)\right\Vert,
\end{equation}
where now the vector is both rotated and projected
\begin{equation}
   \label{eq:diffpredictediter}
   {d^P}_j(\sigma_i) = \sum_{k=1}^{N_{\rm rep}}
   d_k(\sigma_i)
   P_{kj}, \qquad j=1,\dots,N_{\rm eig}\, .
\end{equation}
The norm of $d^P$ is  only  approximately equal to that of the
starting vector of differences $d$: $\left\Vert d^P(\sigma_1)\right\Vert\approx 
\left\Vert {d}(\sigma_1)\right\Vert$.
 However, it  
is easy to see  that
this provides the linear combination of
replicas which 
minimizes the
difference   in absolute value between the prior and final
covariance matrix for given number of eigenvectors.
As the difference decreases  monotonically
as $N_{\rm eig}$ increases, the value of $N_{\rm eig}$ 
can be tuned to any desired accuracy goal, with the exact equality 
Eq.~(\ref{rotation}) achieved when  $N_{\rm eig}=N_{\rm eig}^{(0)}$.
Note that, of course, the optimization step
can be performed also starting with a
symmetric Hessian, rather than Monte Carlo, prior.
In such case, the index $k$
runs over Hessian eigenvectors,  Eq.~(\ref{covmat}) is replaced by
$\textrm{cov}(Q) = XX^t$,  and the rest of the procedure is unchanged.

An interesting feature of this SVD+PCA method is that the matrix $V$
(and thus also the principal matrix  $P$) in Eq.~(\ref{eq:diffpredicted}) does not
depend on the value of the PDF factorization scale $Q$:
the scale dependence is thus entirely given by the
DGLAP evolution equation satisfied by the original Monte Carlo
replicas.
The result of the SVD thus does not depend on the scale at
which it is performed. Of course, the subsequent PCA projection may
depend on scale if there are level crossings, but this is clearly a
minor effect if a large enough number of principal components is
retained. Because of this property, the SVD+PCA methodology can be
used for the efficient construction~\cite{Butterworth:2015oua} of a Hessian representation of
combined PDF sets, even when the sets which enter the combination
satisfy somewhat different evolution equations, {\it e.g.}, because of
different choices in parameters such as the heavy quark masses, or in the specific
solution of the DGLAP equations.

\subsection{The SM-PDF method}
\label{sec:sm-pdf}

In the  SM-PDF method, this same SVD+PCA optimization is performed, but
now with the goal of achieving a given accuracy goal not for the full
prior
PDF set in the complete range of $x$ and $Q^2$,
but rather for the aspects of it which are relevant for the
determination of a given input set of cross-sections, and in such a
way that all the information which is not immediately used is
stored and can be {\it a posteriori} recovered either in
part or fully, {\it e.g.} if one wishes
to add further observables to the input list.

This requires supplementing the SVD+PCA methodology of Ref.~\cite{Carrazza:2015aoa}
with three additional features:
a measure of the accuracy goal; a way of singling out the relevant
part of the covariance matrix; and a way of keeping the information on
the rest of the covariance matrix in such a way that if needed the
full covariance matrix can be recovered at a later stage.

The main input to the algorithm is the set of $N_{\sigma}$
observables which we want to reproduce,
$\{\sigma_i\}$, with  $i=1,\ldots  N_{\sigma}$.
Theoretical predictions for the cross-sections
$\{\sigma_i\}$ are computed using a prior 
PDF set, which we assume for definiteness to be given as a Monte Carlo,
though the method works with obvious modifications also if the
starting PDFs are given in Hessian form.
The goal of the SM-PDF methodology is to evaluate  the
PDF uncertainties   $s_{\sigma_i}$, Eq.~(\ref{eq:obsstd}),
in terms of a reduced number of Hessian eigenvectors,
\begin{equation}
   \label{eq:obsstd2}
   \tilde{s}_{\sigma_i} = \left(\sum_{n=1}^{N_{\rm eig}}\left(\widetilde{\sigma}_i^{(n)} -
   \widetilde{\sigma}_i^{(0)}\right)^2\right)^\frac{1}{2} \ ,
   \end{equation}
with the number $N_{\rm eig}$ being as small as possible within a given
accuracy.
We thus define a measure $T_R$ of the accuracy goal 
(tolerance) by the condition
\begin{align}
   \label{eq:tolerance}
T<T_R;\qquad
  T \equiv \max_{i\in (1,N_{\sigma})} \Bigg| 1 - \frac{\tilde{s}_{\sigma_i}}{s_{\sigma_i}}\Bigg| 
\end{align}
in other words, $T_R$ is the maximum relative difference which is 
allowed between the original
and reduced PDF uncertainties, $\tilde{s}_{\sigma_i}$ and
$s_{\sigma_i}$ respectively, for all the observables
$\{\sigma_i\}$.

In order to determine the
part of the covariance matrix relevant for the
description of the input observables $\{\sigma_i\}$, we define
the correlation function
\be
\label{eq:corr_mc}
\rho\lp x_i,Q, \alpha, \sigma_i\rp \equiv \frac{N_{\rm rep}}{N_{\rm
    rep}-1} \lp \frac{\la X(Q)_{lk}  d_k(\sigma_i)\ra_{\rm rep}- \la
  X(Q_{\sigma_i})_{lk}\ra_{\rm rep}  \la d_k(\sigma_i)\ra_{\rm rep}}{s^{\rm
    PDF}_\alpha(x_i,Q)  s_{\sigma_i} } \rp \, , 
\ee
where the matrix of PDF differences $X(Q)$ and the grid index $l=N_x(\alpha-1)+i$
 have been defined in
Eq.~(\ref{eq:Xmat}); $s_{\alpha}^{\rm PDF}(x_i,Q)$ is
the standard deviation of the PDFs in the prior Monte Carlo representation,
given by the usual expression
\begin{equation}
   \label{eq:errPDF}
   s_{\alpha}^{\rm PDF}(x_i,Q) = \left(   \frac{1}{N_{\rm rep}-1}\sum_{k=1}^{N_{\rm
   rep}} \lc  f_{\alpha}^{(k)}(x_i,Q)-\la f_{\alpha}(x_i,Q)\ra \rc \right)^{\frac{1}{2}}  \, ,
\end{equation}
and $s_{\sigma_i}$, the standard deviation of the $i$-th observable $\sigma_i$,  is
given by Eq.~(\ref{eq:obsstd}).
The function Eq.~(\ref{eq:corr_mc})
measures the correlation between the observables $\sigma_i$ and the
$l$-th PDF value ({\it i.e.}  $f_{\alpha}(x_{i},Q)$, with
$l=N_x(\alpha-1)+i$).

\begin{figure}[t]
  \vspace{-0.7cm}
\begin{center}
  \includegraphics[width=0.85\textwidth]{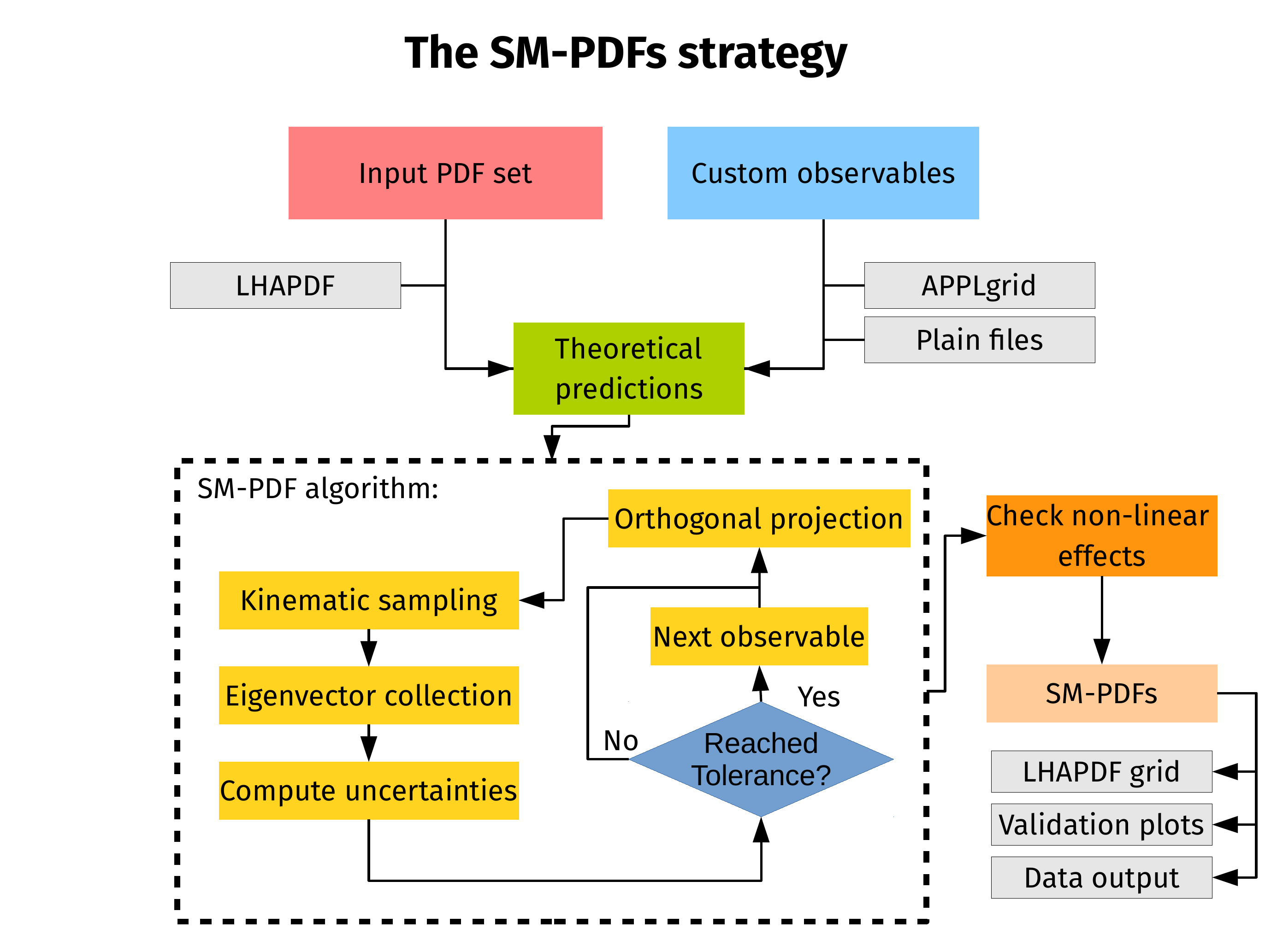}
\end{center}
\vspace{-0.3cm}
\caption{\small Schematic representation of the SM-PDF strategy.\label{fig:scheme}}
\end{figure}

The basic idea of the SM-PDF construction is to apply the SVD
to the subset of the covariance matrix which is most correlated to the
specific 
observables that one wishes to reproduce, through a procedure such
that
information is never discarded, so observables
can be added one at a time, or at a later stage. This goal is achieved
through  an iterative procedure
schematically represented in Fig.~\ref{fig:scheme}, which
we now describe in detail.

The iteration loop (contained in the dashed
box in Figure~\ref{fig:scheme}) is labeled by an iteration
index $j$, such that at each iteration an extra eigenvector is
added, thereby increasing the accuracy.
If the accuracy goal is
achieved for all observables
after  $j$ iterations, then the final reduced Hessian set contains
$N_{\rm eig}=j$ eigenvectors as  error sets.
These are delivered as a new principal matrix
$P$, which provides the expansion coefficients of the eigenvectors over
the replica basis: namely, $P_{kj}$ is the component of the
$j$-th eigenvector in terms of the $k$-th replica. They thus
replace the principal matrix of the previous PCA procedure as a final output
of the procedure, and can be used in exactly the same way.

To set off the iterative procedure, we select one
of the observables we wish to reproduce from the list, $\sigma_1$, 
and compute the correlation coefficient
$\rho\lp x_i,Q, \alpha, \sigma_1\rp$ for all grid points
$(x_i,\alpha)$ and for a
suitable choice of scale $Q$. We then identify the subset $\Xi$ of grid points for which  $\rho$
exceeds some threshold value:
\begin{equation}
  \Xi=\left\{ (x_i,\alpha):
  \rho\lp x_i,Q_{\sigma_1}, \alpha, \sigma_1\rp\ge
   t  \rho_{\max}  \right\} \, .
\label{eq:corrinterval}
\end{equation}
The threshold value is expressed as a fraction $0<t<1$ times
the maximum value $\rho_{\max}$ that the correlation coefficient takes
over the whole
grid, thereby making the criterion independent of the absolute scale
of the correlation.
The choice of scale $Q$ and threshold parameter $t$ should be taken
as tunable settings of the procedure, and will be
discussed in Sect.~\ref{sec:validation} below. For the time being it
suffices to say that $Q$ should be of the order of the typical scale
of the observable (for example, the average value of the factorization
scale).

We  then
 construct a reduced sampling matrix $X_{\Xi}$, defined
as in
Eq.~(\ref{eq:Xmat}),  but now only including 
points in the
$\{x_i, \alpha\}$ space which are in the subset $\Xi$.
We  perform the SVD  of the reduced matrix
\begin{equation}
  \label{eq:svdxi}
X_\Xi=USV^t\ ,  
\end{equation}
and we  only keep the largest principal component,
{\it i.e.}  one single largest eigenvector, which is specified by the
coefficients   of its expansion over the replica basis, namely,  
assuming that the singular values are ordered, by
the first  row of the  $V$ matrix. We thus start filling our output principal
matrix  $P$ by letting
\begin{equation}
   \label{eq:principalvector}
   P_{kj} = V_{k1}^{(j)} \,, \quad j=1 \,, \quad k=1,\ldots,N_{\rm rep} \, .
\end{equation}
Note that $j$ on the left-hand side labels the eigenvector ($P_{kj}$
provides expansion coefficients for the $j$-the eigenvector) and on
the right-hand side it labels the iteration ($V_{k1}^{(j)}$ is the
first row of the $V$-matrix at the $j$-th iteration), which we can
identify because, as mentioned, at each iteration we will add an
eigenvector.
The remaining  eigenvectors of the
principal matrix span the  linear
subspace orthogonal to $P$, and we assign them to a  residual matrix
$R$:
\begin{alignat}{2}
   \label{eq:residual}
   & R_{km}^{(j)} = V_{k(m+1)}^{(j)} \, \qquad j=1 \, , \qquad &&
   m=1,\ldots,N_{\rm rep}-1
   \,, \qquad k=1,\ldots,N_{\rm rep} \, .
\end{alignat}

At the first iteration, when there is only one eigenvector, the
principal matrix $P$ has just one row, and it  coincides with the
principal component of $V$. So far, the procedure is identical to that
of the SVD+PCA method, and we can thus use again
Eq.(\ref{eq:errfinal}) to compute uncertainties on observables, check
whether the condition Eq.~(\ref{eq:tolerance}) is met, and if it is
not add more eigenvectors. 
The  procedure works in such a way that each time a new eigenvector is
selected, exactly the same steps are repeated in the subspace
orthogonal to that of the previously selected eigenvectors, thereby
ensuring that information is never discarded. This is
achieved by a projection method.

Specifically, we project the
 matrix $X$ and the vector of observable
      differences $\{d_k(\sigma_i)\}$ on the
orthogonal subspace of $P$, namely, the space orthogonal to that
spanned by the eigenvectors which have already been selected (as many
as the number of previous iterations).
The projections are performed by respectively 
replacing $d(\sigma_i)$ and $X$ by
\begin{align}
	\label{eq:iterstart}
		d^R(\sigma_i) &=d(\sigma_i) 
		R^{(j-1)} \, ,\\
		 X^R &=X  R^{(j-1)} \, ,
\end{align}
where the first iteration of the residual matrix $R^{(1)}$ has been defined
in Eq.~(\ref{eq:residual}).

After the projection, we  proceed as in the first iteration. We
first determine again the  subset $\Xi$,
Eq.~(\ref{eq:corrinterval}), of the projected sampling matrix $
X^R$, thereby obtaining a new sampling matrix $X^R_\Xi$: this is possible
because everything is
expressed as a linear combination of replicas anyway.  
Once the new matrix $X^R_\Xi$ has been constructed, the procedure is
restarted from Eq.~(\ref{eq:svdxi}),
leading to a new matrix $V^R$.
The number of columns of the projected matrix  $X^R_\Xi$ (and therefore of
$V^R$) is $N_{\rm rep}-(j-1)$, which is the 
dimension of the subspace of the linear combinations not yet selected
by the algorithm (that is, $N_{\rm rep}-1$ for $j=2$, and so on).
We
can now go back to Eq.~(\ref{eq:principalvector}) and proceed as in
the previous case, but with the projected matrices: we add another
row to the matrix of coefficients to the principal matrix by picking the largest
eigenvector of the projected matrix, and determining again the orthogonal
subspace. 

At the $j$-th iteration, this procedure gives
\begin{align}
   \label{eq:principalvectorJ}
   {P^R}^{(j)}_{k} &= {V^R}_{k1}^{(j)} \,, \quad
   k=1,\ldots,N_{\rm rep}-(j-1) \, ,\\
    {R^R}_{km}^{(j)} &={V^R}_{k(m+1)}^{(j)} \,, \quad
   m=1,\ldots,N_{\rm rep}-j
   \,, \quad k=1,\ldots,N_{\rm rep}-(j-1) \, .
\end{align}
which respectively generalize Eqs.~(\ref{eq:principalvector}) and
(\ref{eq:residual}) for $j\geq 1 $. 
The projected row of coefficients 
$P^R$ Eq.~(\ref{eq:principalvectorJ}) can be used to determine the
corresponding unprojected row of coefficients of the principal matrix
and of the residual matrix
by using the
projection $R$ matrix in reverse, {\it i.e.}, at the $j$-th iteration
\begin{align}
	\label{eq:iterend}
		P^{(j)}_{kh} &=\sum_{k'} R^{(j-1)}_{k k'}
                {P^R}^{(j)}_{k'h} \, ,\\
		R^{(j)}_{kh} &=\sum_{k'}  R^{(j-1)}_{kk'}  {R^R}^{(j)}_{k'h}\, .
\end{align}
We thus end up with a principal matrix which has been filled with a further
eigenvector, and a new residual matrix and thus a new projection.

In summary, at each iteration we first project
onto the residual subspace, Eq.~(\ref{eq:iterstart}), then pick the largest eigenvector in the
subspace, Eq.~(\ref{eq:principalvectorJ}), then re-express results in the starting space
of replicas, Eq.~(\ref{eq:iterend}), so  $P$ is always the first row
of $V$ in each subspace, and  
Eqs.~(\ref{eq:diffpredictediter}-\ref{eq:errfinal}) remain valid as
the  $P$ matrix is gradually filled. Determining the correlation
and thus $\Xi$ after projection ensures that only the correlations
with previously unselected linear combinations are kept. The fact that
we are always working in the orthogonal subspace  implies
that  
the  agreement for the observables $\sigma_i$ which had
already been included can  only be improved and not deteriorated by
subsequent iterations. It follows that we can always just check the
tolerance condition on one observable at a time.
The procedure is thus unchanged regardless of whether we are
adding a new observable or not. In any case, 
the  subset $\Xi$ Eq.~(\ref{eq:corrinterval}) 
is always determined by only one observable, namely, the one
that failed to satisfy the tolerance condition at the previous
iteration.
The procedure is iterated until the condition is satisfied for all
observables $\{\sigma_i\}$ in the input list.
The number of iterations $j$ until convergence
defines the final number of eigenvectors
$N_{\rm eig}$.

The output of the algorithm is the final 
$N_{\rm rep}\times N_{\rm   eig}$ principal matrix $P$, which can be
 used to compute uncertainties on observables using
 Eqs.~(\ref{eq:errfinal}-\ref{eq:diffpredictediter}). 
However, for the
 final delivery we wish to obtain a set of Hessian eigenvectors. These
can be obtained by performing the linear transformation given by $P$
(a rotation and a projection)
in the space of PDFs. The $X$ matrix
Eq.~(\ref{eq:Xmat}) then becomes
\begin{equation}
   \label{eq:newX}
   \widetilde{X} \equiv \sqrt{\frac{1}{N_{\rm rep}-1}}X  P \, ,
\end{equation}
 so, 
substituting in Eq.~(\ref{eq:Xmat}), the 
final $N_{\rm eig}$ eigenvectors are found to be given by
\begin{equation}
    \widetilde{f}_{\alpha}^{(k)}(x_i,Q)= f_{\alpha}^{(0)}(x_i,Q) +
  \widetilde{X}_{lk}(Q)\,, \quad
  k=1,\ldots,N_{\rm eig}\, .
  \label{eq:newhessian}
  \end{equation}
  This is the same result as with the SVD+PCA  algorithm of Sect.~\ref{sec:svd-pca}, but
  now generally with a smaller number of eigenvectors, namely, those
  which are necessary to describe the subset of the covariance matrix
  which is correlated to the  input
  set of observables.
   
\subsection{SM-PDF usage and optimization}
\label{sec:optimisation}

Upon delivery of
the final PDF set, any observable is computed 
in terms of the resulting
Hessian representation  Eq.~(\ref{eq:newhessian}). 
As in the case of the original SVD+PCA methodology, the final result  
Eq.~(\ref{eq:newhessian}) determines the PDFs for all $x$ and $Q$. Indeed,
Eq.~(\ref{eq:newhessian}) determines the  SM-PDF Hessian
eigenvectors as linear combinations of replicas, and thus for all
values of $x$ and $Q$ for which the original replicas were defined.

Note however that 
in the procedure of Sect.~\ref{sec:sm-pdf}, in order to test for the
tolerance criterion  observables have been
computed using
Eqs.~(\ref{eq:errfinal}-\ref{eq:diffpredictediter}). This is
equivalent to using the PDFs 
Eq.~(\ref{eq:newX})  by  standard linear
error propagation, but it differs from it by nonlinear 
terms, specifically for hadron collider processes in which observables
are quadratic in the PDFs.
Even though nonlinear corrections are expected to be small, in
principle it could be that the tolerance criterion is no longer
satisfied if Eq.~(\ref{eq:newX}) is used instead. 

We explicitly check for this, and if it is the case for all observables
$\sigma_i$ such that the recomputed tolerance criterion is not satisfied, 
we restart the iteration but now replacing 
the tolerance with a new value
$T_{R,i}^{\rm (new)}$ given by
\begin{equation}
	\label{eq:newtol}
	T_{R,i}^{\rm (new)} \equiv T_R -  \lp T_i - T_i^{\rm (lin)} \rp \, ,
\end{equation}
where $T_i^{\rm (lin)}$ is the value of the
tolerance that is obtained within the linear
approximation, by computing Eq.~(\ref{eq:tolerance}) with
Eq.~(\ref{eq:errfinal}).
Iterating until the criterion with  the new tolerances
Eq.~(\ref{eq:newtol}) is met
will  be sufficient  to ensure that the tolerance criterion is
satisfied when using the new PDFs, provided the difference between the
linear and exact estimate of $T_i$ is mostly due to  the larger
eigenvectors that were selected first, and remains approximately
constant upon addition of  smaller eigenvectors in order to correct for this.

In practice, the difference between the linear estimation of
the PDF uncertainty and the exact result is generally
 small, and does not a change the result for target
tolerances $T_R$ of $5\%$ or bigger.
This effect can be more important for observables
affected by substantial PDF uncertainties, or for processes
which depend on a large number of partonic channels (especially
when new channels open up at NLO or NNLO). It is
however not an issue for most practical applications.

Note that  this final optimization step may become
extremely
time consuming if fast grid tools are not available.
In view of this, it is possible to disable this check.
However, fast
interfaces can be obtained for any NLO QCD cross-section
with arbitrary final-state cuts using
the {\tt aMCfast} interface~\cite{Bertone:2014zva} to
{\tt Madgraph5\_aMC@NLO}~\cite{Alwall:2014hca}.

The SM-PDF construction can be generally performed at any perturbative
order, and specifically starting with an NLO or an NNLO PDF set. The
perturbative order enters both in the choice of starting PDF set, and
in the computation of the list of observables $\{\sigma_i\}$,
specifically used for the determination of the correlation function $\rho$
Eq.~(\ref{eq:corr_mc}). Because the NNLO-NLO $K$ factors are usually
moderate, for most applications it may be sufficient to compute
$\rho$ using NLO theory even when using NNLO PDFs throughout. An
obvious exception is the case in which the user is explicitly
interested in studying the changes in PDFs when going from NLO to
NNLO. 

A final issue is whether results depend on the order in which the
observables are included, and specifically on the choice of the
observable $\sigma_1$ used to start the iteration.
Indeed, the
eigenvectors selected for a specific  iteration depend on the 
subspace spanned by the previous eigenvectors, and consequently  a different
ordering will indeed change the particular linear combinations that
are selected. 
However this does not significantly affect the total number of
eigenvectors needed, because the optimal subspace of linear
combinations  required to describe all observables with a given accuracy
remains the same regardless of the order they are presented. We have
verified that this is indeed the case, though we observed small
fluctuations by one or two units in the final number of eigenvectors due to the
discontinuous nature of the tolerance criteria
Eq.~(\ref{eq:tolerance}).

\section{Results and validation}
\label{sec:validation}

We now present the
validation of the  SM-PDF algorithm described
in the previous section.
Using this methodology, we have constructed four
specialized minimal PDF sets for different
representative cases of direct
phenomenological relevance at the LHC:
\begin{enumerate}
\item Higgs physics,
\item Top quark pair production physics,
\item Electroweak gauge boson production physics,
  \item The combination of all processes included in (1), (2) and (3).
\end{enumerate}
These examples have been chosen since, for each SM-PDF,
there is a strong  case for the use of optimized
PDF sets with a greatly reduced number of eigenvectors.
For instance, these SM-PDFs could be of interest for studies
of the Higgs
Cross-Section Working Group~\cite{Dittmaier:2011ti} (case 1), the LHC
Top Working Group (case 2), and
the LHC Electroweak Working Group (case 3), respectively.
As an example,
the SM-PDFs for $W,Z$ production could be relevant for the determination
of the $W$ boson mass~\cite{Bozzi:2011ww,Bozzi:2015zja,Bozzi:2015hha},
which is a extremely CPU-time consuming task.

In this section,
we will  first define the PDF priors and
LHC cross-sections
that have been used to construct the  SM-PDF sets listed
above, then validate the performance of the algorithm
using a variety of figures of merit.

\subsection{Input PDFs and cross-sections}
\label{sec:priors}

In order to validate the  SM-PDF methodology, we have used three different
prior PDF sets, all of them in the Monte Carlo representation:
\begin{enumerate}
\item The NNPDF3.0
  NLO set ~\cite{Ball:2014uwa}  with $N_{\rm rep}=1000$ replicas,
\item The MMHT14 NLO set~\cite{Harland-Lang:2014zoa}
  with $N_{\rm rep}=1000$ replicas, obtained from
  the native Hessian representation using the Watt-Thorne method~\cite{Watt:2012tq}, and
\item The PDF4LHC 2015 NLO prior set~\cite{Butterworth:2015oua},
 with $N_{\rm rep}=900$ replicas, built  from the combination
  of 300 replicas from each of the CT14, MMHT14 and
  NNPDF3.0 NLO sets.
  This set is denoted by MC900 in the following.
\end{enumerate}
These three choices are representative enough for the validation of
our methodology; they show  that the procedure works regardless of the choice of
input PDF set. As already mentiond in Sect.~\ref{sec:optimisation}
the SM-PDF methodology can be applied equally to NLO or NNLO PDFs, and
NLO PDFs are chosen here purely for the sake of illustration. Indeed,
in Appendix~\ref{sec-appendix} we provde an example in which NNLO PDFs
are used.

\begin{table}[t]
\footnotesize
\begin{centering}
  \begin{tabular}{|c|c|c|c|c|c|}
    \hline
    \multicolumn{6}{|c|}{Input cross-sections for
      SM-PDFs for Higgs physics} \\
    \hline 
\hline 
process & distribution & grid name &  $N_{{\rm bins}}$ & range & kin.
cuts\tabularnewline
\hline 
\hline 
$gg\to h$ & incl xsec &{\tt ggh\_13tev}  & 1 & - & -\tabularnewline
 &  $d\sigma/dp_t^h$  &{\tt ggh\_pt\_13tev}  & 10 & {[}0,200{]} GeV & -\tabularnewline
 & $d\sigma/dy^h$ & {\tt ggh\_y\_13tev}  & 10 & {[}-2.5,2.5{]} & -\tabularnewline
\hline 
VBF $hjj$ & incl xsec & {\tt vbfh\_13tev}  & 1 & - & -\tabularnewline
&  $d\sigma/dp_t^h$  &{\tt vbfh\_pt\_13tev}  & 5 & {[}0,200{]} GeV & -\tabularnewline
& $d\sigma/dy^h$ & {\tt vbfh\_y\_13tev}  & 5 & {[}-2.5,2.5{]} & -\tabularnewline
\hline 
$hW$& incl xsec & {\tt hw\_13tev}  & 1 & - & $p_{T}(l)\geq10$ GeV, $|\eta^{l}|\leq2.5$\tabularnewline
&  $d\sigma/dp_t^h$  &{\tt hw\_pt\_13tev}  & 10 & {[}0,200{]} GeV & $p_{T}(l)\geq10$ GeV, $|\eta^{l}|\leq2.5$\tabularnewline
 & $d\sigma/dy^h$ & {\tt hw\_y\_13tev}  & 10 & {[}-2.5,2.5{]} & $p_{T}(l)\geq10$ GeV, $|\eta^{l}|\leq2.5$\tabularnewline
\hline 
$hZ$ & incl xsec & {\tt hz\_13tev}  & 1 & - & $p_{T}(l)\geq10$ GeV, $|\eta^{l}|\leq2.5$\tabularnewline
&  $d\sigma/dp_t^h$  &{\tt hz\_pt\_13tev}  & 10 & {[}0,200{]} GeV & $p_{T}(l)\geq10$ GeV, $|\eta^{l}|\leq2.5$\tabularnewline
& $d\sigma/dy^h$ & {\tt hz\_y\_13tev}  & 10 & {[}-2.5,2.5{]} & $p_{T}(l)\geq10$ GeV, $|\eta^{l}|\leq2.5$\tabularnewline
\hline
$ht\bar{t}$ & incl xsec & {\tt httbar\_13tev}  & 1 & - & -\tabularnewline
&  $d\sigma/dp_t^h$  &{\tt httbar\_pt\_13tev}  & 10 & {[}0,200{]} GeV & -\tabularnewline
 & $d\sigma/dy^h$ & {\tt httbar\_y\_13tev}  & 10 & {[}-2.5,2.5{]} & -\tabularnewline
\hline
\end{tabular}
\par\end{centering}
\caption{\small LHC processes and the corresponding
  differential distributions that have been
  used as input in the construction of the
  SM-PDFs
  dedicated to Higgs physics.
  In each case we also provide the {\tt APPLgrid} grid name, the range
  spanned by each distribution, the number of
  bins $N_{\rm bins}$, and the kinematical cuts applied to the final-state
  particles.
  For associated production with vector bosons, $hW$ and $hZ$, we impose
  basic acceptance cuts on the charged leptons from the weak boson decays.
  All processes have been computed for the LHC 13 TeV.
  \label{tab:processes_H}}
\end{table}

In order to compute the theoretical predictions for all input
PDF sets and as many cross-sections as possible,
we have generated a large number of
dedicated {\tt APPLgrid} grids~\cite{Carli:2010rw}
using the {\tt aMCfast}~\cite{Bertone:2014zva} interface
to  {\tt MadGraph5\_aMC@NLO}~\cite{Alwall:2014hca}.
Cross-sections and differential distributions have been computed
for the LHC Run II kinematics,
with a center-of-mass energy of
$\sqrt{s}=13$ TeV.
In particular we have generated fast NLO grids
for the following processes:
\begin{itemize}

\item Higgs production: total cross-sections and rapidity and $p_T$
  differential distributions for 
  gluon-fusion, vector-boson fusion, associated production with $W$ and
  $Z$ bosons and associated production with top quark pairs.
  No Higgs decays are included, since we are only interested
  in the production dynamics.

\item Top quark pair production: total cross-section,
   $p_t$ and rapidity distributions of the top and the anti-top quarks,
  and invariant mass $m_{t\bar{t}}$,  $p_t$, and
  rapidity distributions of the
  $t\bar{t}$ system.

\item Electroweak gauge
  boson production.
  For $Z$ production: total cross-section,  $p_T$
  and rapidity distributions of the two charged leptons and of the $Z$
  boson, and  $p_T$ and invariant mass distribution of the dilepton pair.
  For $W$ production: total cross-section, $p_T$
  and rapidity distributions of the charged lepton and of the $W$ boson,
  missing $E_T$  and transverse mass $m_T$ distribution.
    For the $W$ and $Z$ processes,
  we apply kinematical cuts to the charged leptons
  from the weak boson decay to reflect the typical
  acceptance constraints of the LHC experiments.

\end{itemize}

  A more detailed description of these
  processes, including binning and
  the kinematical cuts applied, is provided in
  Tables~\ref{tab:processes_H}--\ref{tab:processes_WZ}.
  We also indicate the names of the (publicly available)
  {\tt APPLgrid} grids generated for the present validation study.
  Producing fast NLO grids for additional processes, or with
  a different binning or set of analysis cuts, is
  straightforward using the {\tt aMC@NLO}/{\tt aMCfast} framework.
  We adopt the default choice of renormalization and factorization
  scales in {\tt aMC@NLO}, namely
  $\mu_F=\mu_R=H_T/2$,
with
\be
H_T\equiv \sum_i \sqrt{p_{T,i}^2+m_i^2} \, ,
\ee
the scalar sum of the
transverse masses of all final state particles at the
matrix-element level.

\begin{table}[t]
\small
\begin{centering}
  \begin{tabular}{|c|c|c|c|c|c|}
    \hline
    \multicolumn{6}{|c|}{Input cross-sections for  SM-PDFs for  $t\bar{t}$ physics} \\
    \hline 
\hline 
process & distribution & grid name &  $N_{{\rm bins}}$ & range & kin.
cuts\tabularnewline
\hline 
\hline 
$t\bar{t}$ & incl xsec &{\tt ttbar\_13tev}  & 1 & - & -\tabularnewline
 & $d\sigma/dp_t^{\bar{t}}$  & {\tt ttbar\_tbarpt\_13tev}  & 10 & {[}40,400{]} GeV & -\tabularnewline
 & $d\sigma/dy^{\bar{t}}$ &{\tt ttbar\_tbary\_13tev}  & 10 & {[}-2.5,2.5{]} & -\tabularnewline
 & $d\sigma/dp_t^{t}$ &{\tt ttbar\_tpt\_13tev}  & 10 & {[}40,400{]} GeV & -\tabularnewline
 & $d\sigma/dy^{t}$&{\tt ttbar\_ty\_13tev}  & 10 & {[}-2.5,2.5{]} & -\tabularnewline
 & $d\sigma/dm^{t\bar{t}}$ &{\tt ttbar\_ttbarinvmass\_13tev}  & 10 & {[}300,1000{]} & -\tabularnewline
 & $d\sigma/dp_t^{t\bar{t}}$  &{\tt ttbar\_ttbarpt\_13tev}  & 10 & {[}20,200{]} & -\tabularnewline
 & $d\sigma/dy^{t\bar{t}}$ &{\tt ttbar\_ttbary\_13tev}  & 12 & {[}-3,3{]} & -\tabularnewline
\hline 
\end{tabular}
\par\end{centering}
\caption{\small Same as Table~\ref{tab:processes_H}
  for the
  SM-PDFs
  dedicated to top-quark pair production physics.
 \label{tab:processes_ttbar}}
\end{table}

\begin{table}[t]
\footnotesize
\begin{centering}
  \begin{tabular}{|c|c|c|c|c|c|}
    \hline
    \multicolumn{6}{|c|}{Input cross-sections for SM-PDFs for electroweak boson production physics} \\
    \hline 
  \hline
  process & distribution & grid name  & $N_{{\rm bins}}$ & range & kin.
cuts\tabularnewline
\hline 
\hline 
$Z$ & incl xsec &{\tt z\_13tev}  & 1 & - & $p_{T}(l)\geq10$ GeV,
$|\eta^{l}|\leq2.5$ \tabularnewline
 &   $d\sigma/dp_t^{l^-}$ &{\tt z\_lmpt\_13tev}  & 10 & {[}0,200{]} GeV & $p_{T}(l)\geq10$
 GeV, $|\eta^{l}|\leq2.5$ \tabularnewline
 & $d\sigma/dy^{l^-}$ &{\tt z\_lmy\_13tev}  & 10 & {[}-2.5,2.5{]} & $p_{T}(l)\geq10$
 GeV, $|\eta^{l}|\leq2.5$ \tabularnewline
 & $d\sigma/dp_t^{l^+}$ &{\tt z\_lppt\_13tev}  & 10 & {[}0,200{]} GeV & $p_{T}(l)\geq10$
 GeV, $|\eta^{l}|\leq2.5$ \tabularnewline
 & $d\sigma/dy^{l^-}$ &{\tt z\_lpy\_13tev}  & 10 & {[}-2.5,2.5{]} & $p_{T}(l)\geq10$ GeV, $|\eta^{l}|\leq2.5$\tabularnewline
 & $d\sigma/dp_t^{z}$ &{\tt z\_zpt\_13tev}  & 10 & {[}0,200{]} GeV & $p_{T}(l)\geq10$ GeV, $|\eta^{l}|\leq2.5$\tabularnewline
 &  $d\sigma/dy^{z}$&{\tt z\_zy\_13tev}  & 5 & {[}-4,4{]} & $p_{T}(l)\geq10$ GeV, $|\eta^{l}|\leq2.5$\tabularnewline
 & $d\sigma/dm^{ll}$&{\tt z\_lplminvmass\_13tev}  & 10 & {[}50,130{]} GeV & $p_{T}(l)\geq10$ GeV, $|\eta^{l}|\leq2.5$\tabularnewline
 & $d\sigma/dp_t^{ll}$ &{\tt z\_lplmpt\_13tev}  & 10 & {[}0,200{]} GeV & $p_{T}(l)\geq10$ GeV, $|\eta^{l}|\leq2.5$\tabularnewline
 \hline
 \hline
$W$ & incl xsec &{\tt w\_13tev}  & 1 & - & $p_{T}(l)\geq10$ GeV,
$|\eta^{l}|\leq2.5$ \tabularnewline
 &   $d\sigma/d\phi$  &{\tt w\_cphi\_13tev}  & 10 & {[}-1,1{]} & $p_{T}(l)\geq10$
 GeV, $|\eta^{l}|\leq2.5$ \tabularnewline
 &  $d\sigma/dE_t^{\rm miss}$ &{\tt w\_etmiss\_13tev}  & 10 & {[}0,200{]} GeV & $p_{T}(l)\geq10$
 GeV, $|\eta^{l}|\leq2.5$ \tabularnewline
 & $d\sigma/dp_t^{l}$ &{\tt w\_lpt\_13tev}  & 10 & {[}0,200{]} GeV & $p_{T}(l)\geq10$
 GeV, $|\eta^{l}|\leq2.5$ \tabularnewline
 &  $d\sigma/dy^{l}$  &{\tt w\_ly\_13tev}  & 10 & {[}-2.5,2.5{]} & $p_{T}(l)\geq10$ GeV, $|\eta^{l}|\leq2.5$\tabularnewline
 & $d\sigma/dm_{t}$  &{\tt w\_mt\_13tev}  & 10 & {[}0,200{]} GeV & $p_{T}(l)\geq10$ GeV, $|\eta^{l}|\leq2.5$\tabularnewline
 & $d\sigma/dp_t^{w}$  &{\tt w\_wpt\_13tev}  & 10 & {[}0,200{]} GeV & $p_{T}(l)\geq10$ GeV, $|\eta^{l}|\leq2.5$\tabularnewline
 & $d\sigma/dy^{w}$  & {\tt w\_wy\_13tev}  & 10 & {[}-4,4{]} & $p_{T}(l)\geq10$ GeV, $|\eta^{l}|\leq2.5$\tabularnewline
\hline 
\end{tabular}
\par\end{centering}
\caption{\small
Same as Table~\ref{tab:processes_H}
  for the
  SM-PDFs
  dedicated to electroweak gauge boson production physics.
  The kinematical cuts are applied to the charged leptons
  from the weak boson decays.
  \label{tab:processes_WZ}}
\end{table}
  
Clearly, some of these cross-sections 
contain overlapping information, so our list is partially redundant.
For instance, if differential distributions are reproduced,
this will be also the case for total inclusive cross-sections.
Similarly, the rapidity distributions of the $W$ and
$Z$ bosons are closely related to the rapidity distributions
of the leptons from their decay, so including both distributions
will lead to a certain degree of redundancy.

This redundancy
can be used to provide non-trivial check of our methodology.
For instance,
we have verified that by beginning
with the total cross-sections, only the most extreme bins of the
differential distributions, which contribute less to the cross
section, might require extra eigenvectors in order to be reproduced to
the desired tolerance.
Conversely, if we begin the algorithm using differential
distributions as input, no additional eigenvectors
are required to describe the corresponding
total cross-sections.

\subsection{Choice of settings}

The SM-PDF method is fully determined by the choice of kinematic
region $\Xi$, Eq.~(\ref{eq:corrinterval}), which in turn is fully
specified by the correlation function and tolerance $T_R$.
The only
tunable parameters are thus the scale $Q$ used for the evaluation of
correlations in Eq.~(\ref{eq:corr_mc}) and the threshold value $t$.
As the choice of scale $Q$, we adopt
the mean value of the factorization scale
$\mu_F$ at which the PDFs are evaluated by the corresponding
{\tt APPLgrid} grids, that is,
 the event-by-event weighted average of
the value of $\mu_F$ used in the calculation of
each specific cross-section or differential
distribution.
 
The only remaining free parameter is then the threshold  $t$, which
specifies
according to  Eq.~(\ref{eq:corrinterval}) which points are included in the reduced
matrix  $X|_{\Xi}$: low values of $t$ lead to the inclusion of a wider region in
phase space, and conversely. Clearly, if $\Xi$ is too
wide,
the reduction will not be very effective
and the ensuing number
of eigenvectors will be large.
On the other hand, if the region $\Xi$
 is too
small, the number of eigenvectors will be small, but it might be lead
to a result which is unstable upon small changes of the input observables.

\begin{figure}[t] 
\begin{center}
\includegraphics[width=0.57\textwidth]{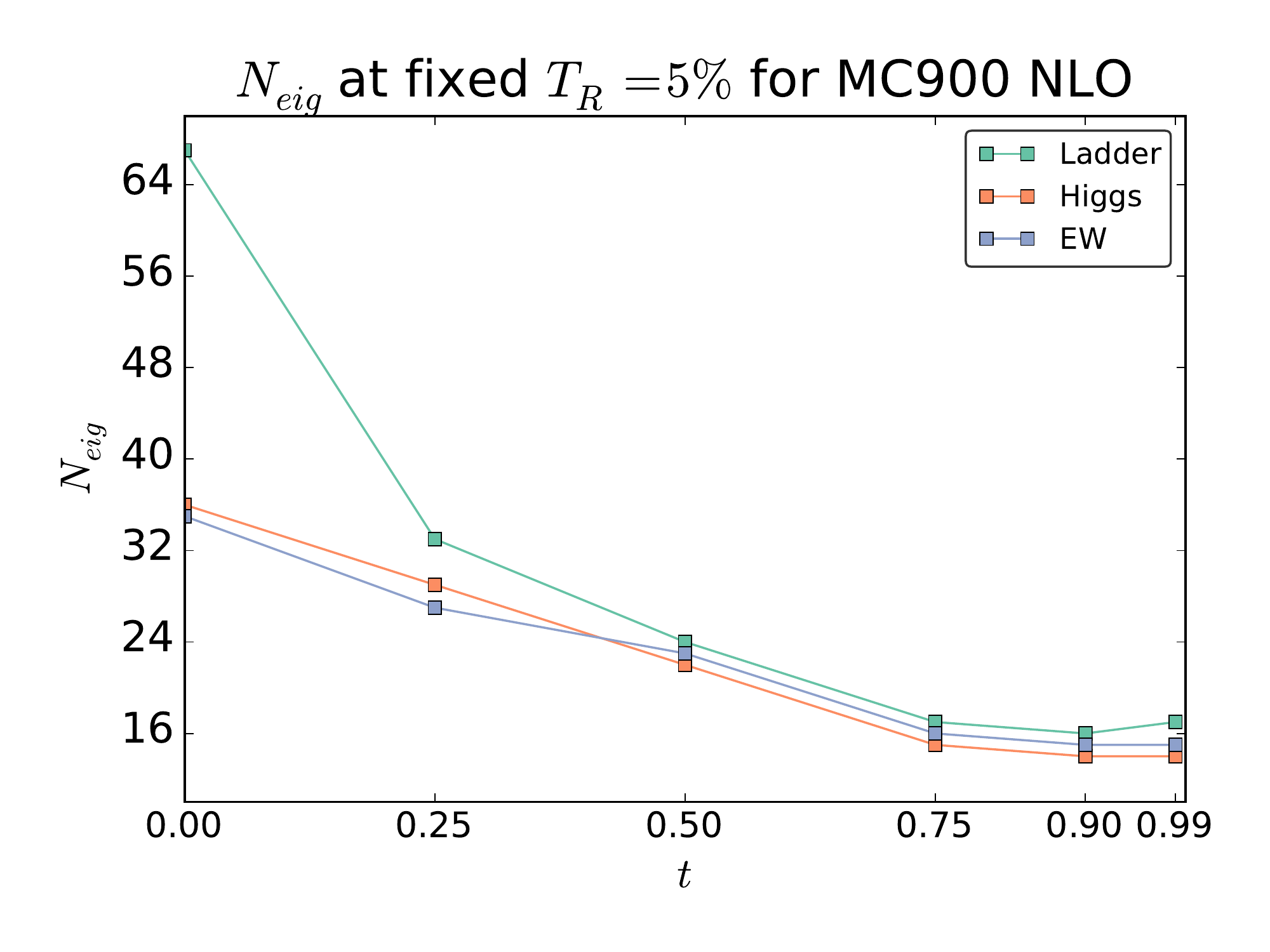}\includegraphics[width=0.42\textwidth]{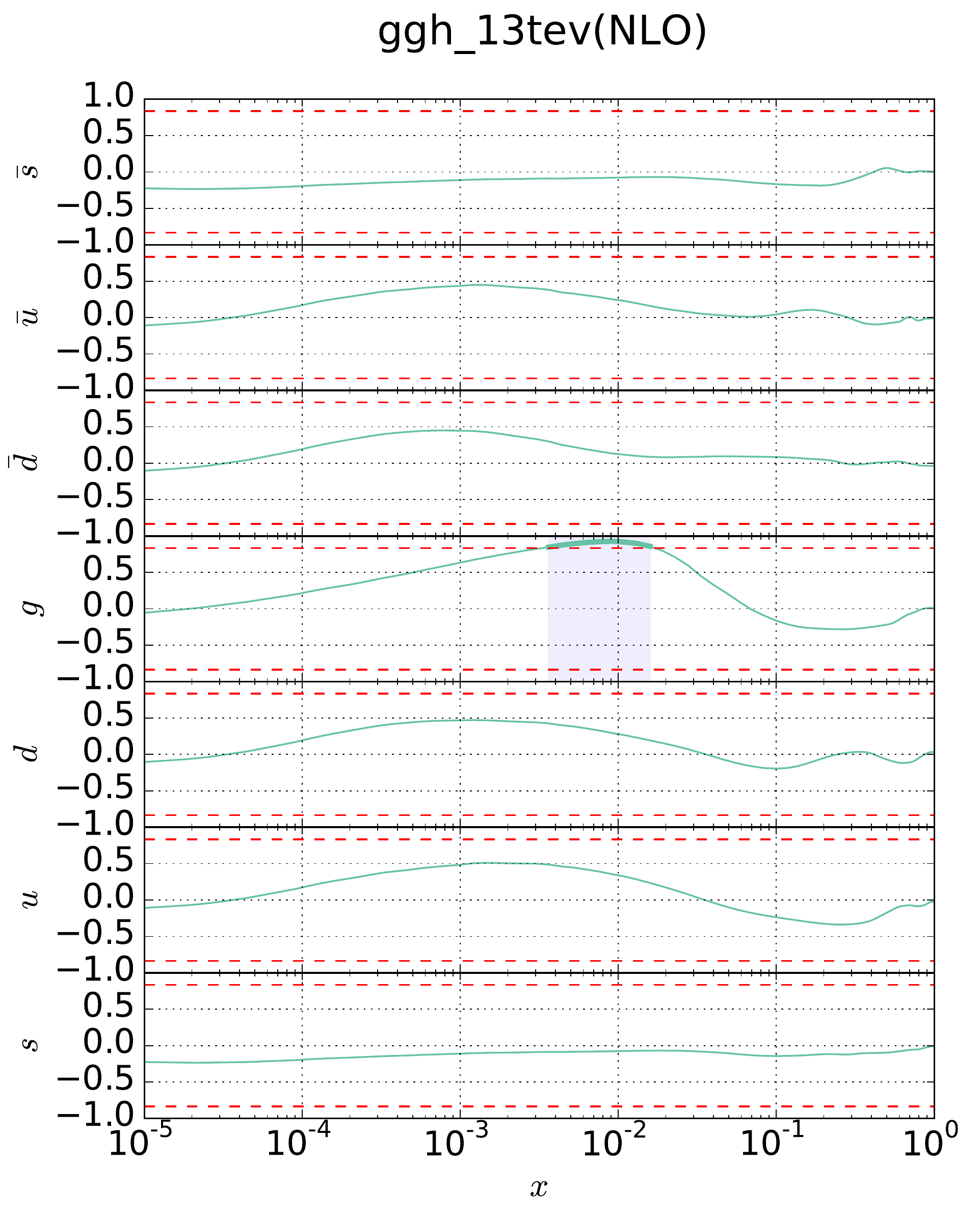}
\end{center}
\vspace{-0.3cm}
\caption{\small (left) Final number of eigenvectors $N_{\rm eig}$
  obtained applying the SM-PDF algorithm to the MC900 NLO PDF set
  with 900 Monte Carlo replicas, as a function of the threshold
  parameter $t$ Eq.~(\ref{eq:corrinterval}) for fixed tolerance $T_R=5\%$.
  We show the results for three choices of input cross-sections:
  Higgs (Table~\ref{tab:processes_H}), electroweak gauge boson production
  (Table~\ref{tab:processes_WZ}),
  and ``ladder'' (all processes in Tables~\ref{tab:processes_H} to~\ref{tab:processes_WZ}).
  (right) Correlation Eq.~(\ref{eq:corr_mc})  between all the PDFs and the total cross-section for Higgs production
  in gluon fusion, as a function of $x$  (solid blue lines).
  The value $\rho = 0.9\rho_{\rm max}$ is shown as a  dashed line and
  the region in which the correlation exceeds the threshold  is shown
  as a shaded band. 
  \label{fig:corrthreshold}}
\end{figure}

In order to determine a suitable value of $t$,
we use the full set of cross-sections listed
in Tables~\ref{tab:processes_H} to~\ref{tab:processes_WZ}.
We will
henceforth refer to this specific set of observables (and the associate
SM-PDF set) as the ``ladder''.
In Fig.~\ref{fig:corrthreshold} (left) we plot
the number of eigenvectors $N_{\rm eig}$ that we obtained
as a function of the parameter $t$ when the  SM-PDF
methodology is applied to the MC900 prior set,
for a fixed tolerance $T_R=5\%$.
We show the results for the Higgs, EW and the ``ladder'' set
of input processes.

As expected, $N_{\rm eig}$ decreases as the value of $t$
is raised, since in this case fewer points in the
$(\alpha,x)$ grid are selected.
While the specific position of the minimum of the $N_{\rm eig}(t)$ curve
depends on the input set of cross-sections,
we see from Fig.~\ref{fig:corrthreshold} that the curve
reaches its minimum around $t\sim 0.9$ for all processes. 
Note that, as discussed at the
end of Sect.~\ref{sec:optimisation}, the value of $N_{\rm eig}(t)$
can fluctuate, typically by one or two units, 
depending on the specific ordering of the input processes.
We therefore choose $t=0.9$: this means that we adopt the smallest
value of $t$ (i.e. the widest kinematic region) compatible with having
the smallest possible number of eigenvectors.

In Fig.~\ref{fig:corrthreshold} (right)
we show the value of the correlation coefficient Eq.~(\ref{eq:corr_mc})
between
the MC900 prior set and the inclusive cross-section for Higgs production
in gluon fusion, as a function of $x$ and for the seven independent
PDF flavors, evaluated at the average scale $Q$ of the grids.
The value of the
correlation $\rho = t\rho_{\rm max}$ corresponding to $t=0.9$ is shown
as a dashed red line in the plots; 
the points for which the correlation coefficient (blue curve)
is larger in modulus than the threshold are shown as a shaded region.

We observe that, for this specific cross-section, the algorithm
in the first iteration will include in the region $\Xi$
Eq.~(\ref{eq:corrinterval}) only the gluon PDF
for $x\simeq 10^{-2}$, which corresponds to
the region that dominates the total cross-section for
Higgs production in gluon fusion.
In Appendix~\ref{sec-appendix-correlations} we provide
additional correlation plots, similar to Fig.~\ref{fig:corrthreshold} (right)
but for other Higgs production channels, as well as
the correlation plots for subsequent iterations, $j\ge 2$, of the
algorithm, illustrating how the selected regions
in the $\lp x,\alpha\rp$ grid vary along the iteration.

\subsection{Results and validation}

We now present the
results of applying the SM-PDF procedure to the PDF sets
and cross-sections described in Sect.~\ref{sec:priors}.
In Table~\ref{tab:neig} we
show the results
for the number of eigenvectors $N_{\rm eig}$ obtained,
for each input
PDF set, using the three different groups
of LHC processes that we consider: Higgs, $t\bar{t}$,
and $W/Z$ production.
In addition, for the Higgs production processes,
we have also studied the results of applying
our methodology to each
of the Higgs production channels individually, as 
summarized in Table~\ref{tab:higgsneig}.
The algorithm has been applied for two different values
of the tolerance  $T_R$, namely 5\% and 10\%.
We also indicate in the bottom row the results
for the  ``ladder'' SM-PDF  (i.e. including all  the above processes.)

\begin{table}[t]
\begin{center}
\begin{tabular}{c||c|c||c|c||c|c}
  \hline
  & \multicolumn{6}{c}{$N_{\rm eig}$} \tabularnewline
    \hline 
\multirow{2}{*}{Process}  & \multicolumn{2}{c||}{MC900} & \multicolumn{2}{c||}{ NNPDF3.0} & \multicolumn{2}{c}{MMHT14}\tabularnewline
\cline{2-7} 
 & $T_R=5\%$ & $T_R=10\%$ & $T_R=5\%$ & $T_R=10\%$ & $T_R=5\%$ & $T_R=10\%$\tabularnewline
\hline 
\hline 
$h$ & 15 & 11 & 13 & 8 & 8 & 7\tabularnewline
$t\bar{t}$ & 4 & 4 & 5 & 4 & 3 & 3\tabularnewline
$W,Z$ & 14 & 11 & 13 & 8 & 10 & 9\tabularnewline
\hline 
\hline 
ladder & 17 & 14 & 18 & 11 & 10 & 10\tabularnewline
\hline 
\end{tabular}
\end{center}
\caption{\small Number of eigenvectors $N_{\rm eig}$ obtained by applying the SM-PDF
  procedure, starting from each of the three
input prior PDF sets,
  to  the three families of processes summarized in Tables
  \ref{tab:processes_H} to~\ref{tab:processes_WZ}: Higgs production,
  $t\bar{t}$ production,
  and $W/Z$ production physics.
  The final row is based on the inclusion of
  all the three families of processes, in the same
  order as they are listed.
  Results are shown for two different values of the tolerance threshold $T_R$,
  5\% and 10\% respectively.
  \label{tab:neig}}
\end{table}

\begin{table}[t]
\begin{center}
  \begin{tabular}{c||c|c||c|c||c|c}
     \hline
  & \multicolumn{6}{c}{$N_{\rm eig}$} \tabularnewline
\hline 
\multirow{2}{*}{Process} & \multicolumn{2}{c||}{MC900} & \multicolumn{2}{c||}{ NNPDF3.0} & \multicolumn{2}{c}{MMHT14}\tabularnewline
\cline{2-7} 
 & $T_R=5\%$ & $T_R=10\%$ & $T_R=5\%$ & $T_R=10\%$ & $T_R=5\%$ & $T_R=10\%$\tabularnewline
\hline 
\hline 
$gg\rightarrow h$ & 4 & 5 & 4 & 4 & 3 & 3\tabularnewline
VBF $hjj$ & 7 & 5 & 10 & 5 & 4 & 3\tabularnewline
$hW$ & 6 & 5 & 6 & 4 & 6 & 3\tabularnewline
$hZ$ & 11 & 7 & 6 & 4 & 8 & 5\tabularnewline
$ht\bar{t}$ & 3 & 2 & 4 & 4 & 3 & 2\tabularnewline
\hline 
\hline 
Total $h$ & 15 & 11 & 13 & 8 & 8 & 7\tabularnewline
\hline 
\end{tabular}
\end{center}
\caption{\small Same as Table~\ref{tab:neig}, now
  for the case where the separate Higgs production channels
  as used as input to the SM-PDF algorithm. \label{tab:higgsneig}}
\end{table}

Several comments on  Table~\ref{tab:neig} are in order.
  \begin{itemize}
  \item Results are reasonably  stable upon a change of  tolerance, with
 differences smaller with the MMHT14 prior, which has smaller
 underlying number of parameters than NNPDF3.0.

  \item The most dramatic reduction in number of eigenvectors is seen
    for the production  of top pairs, or Higgs in gluon fusion,
    where only $N_{\rm eig} \simeq 4$ eigenvectors are needed.
    This can be understood as a consequence of the fact that in both
    cases the dominant contribution to the
    cross-section arises from the gluon distribution
    in a narrow region of $x$.
    
  \item Total
    cross-sections and
    differential distributions for all the Higgs production modes
    can be reproduced, in the case of the MC900
    prior, with  11 to 15 eigenvectors (depending
    on the choice of tolerance $T_R$). 

  \item The number of eigenvectors required  is largest for
    the Higgs and
    the  $W/Z$ family of processes, as one would expect given
    that in both cases several PDFs in a wide kinematic range are required.
    
    \item All the processes that we are including can be described 
      with a SM-PDF set, the
      ``ladder'', which includes about the same number of eigenvectors as
  needed for the  Higgs or  for the
  the Drell-Yan and $W/Z$ family of processes.
  This  ``ladder'' SM-PDF, with only
  $N_{\rm eig}\simeq$ 15 eigenvectors, can be used reliably for
   a large number of LHC cross-sections, including those not
  included in its construction.
    \end{itemize}

\begin{figure}[H]
 \begin{center}
\includegraphics[width=0.95\textwidth]{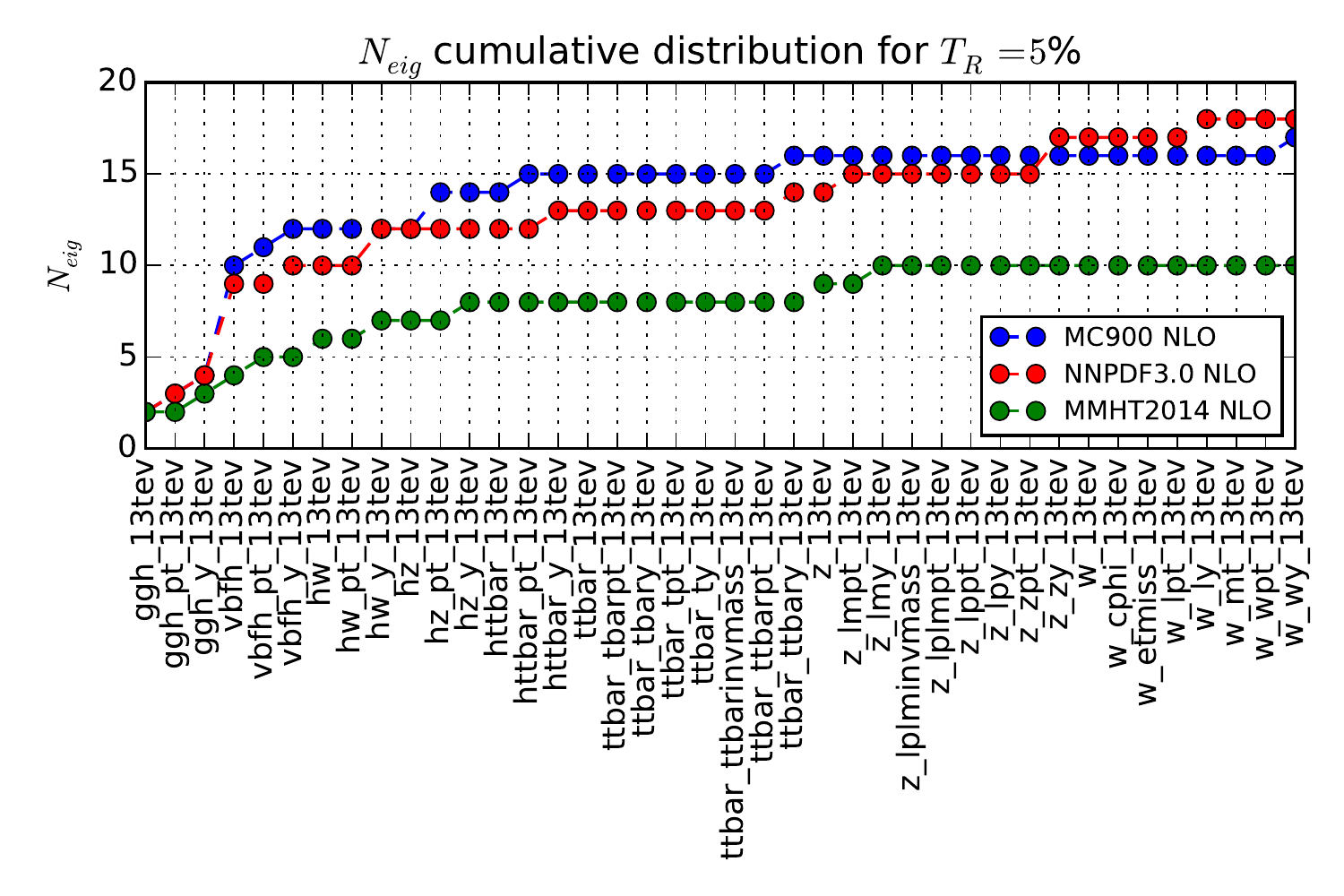} 
\end{center}
\vspace{-1.2cm} \caption{\small Total number of 
  eigenvectors $N_{\rm eig}$ required
  by the SM-PDF algorithm
  to describe a sequentially increasing
  number of input  cross-sections and distributions, for a 
  tolerance $T_R=5$\%.
  Results are presented for the three prior PDF sets, namely
  MC900, NNPDF3.0 and MMHT14.\label{fig:neigevol}}
\end{figure}

  Next, in Fig.~\ref{fig:neigevol} we show the total 
  number of eigenvectors $N_{\rm eig}$ which are required, for a tolerance
  of $T_R=5$\%, as more and more processes are sequentially included, until 
 the complete list of processes in Tables~\ref{tab:processes_H}
 to~\ref{tab:processes_WZ} has been exhausted.
 This plot demonstrates
the robustness and flexibility of the SM-PDF algorithm, in that it
shows how more processes can be added without information loss to a
reduced PDF set, thereby allowing for a study of the information
brought in by each process.
 In   Fig.~\ref{fig:neigevol} results are presented for the three input PDF sets,
   MC900, NNPDF3.0 and MMHT14.
As already seen   
in Table~\ref{tab:neig}, a smaller number of eigenvectors
is required in order to describe the MMHT14 set, which has a smaller
underlying number of parameters than the NNPDF3.0 set; the combined 
MC900 set requires roughly the same number of eigenvectors as
NNPDF3.0, which is contained in it.
Inspection of
Fig.~\ref{fig:neigevol} indicates
which processes bring in new information
in comparison to those already included.
For instance, the fact that the number of eigenvectors is unchanged when adding all
the observables related to top quark
pair production shows that
SM-PDFs based on Higgs processes also describe top
production.

\begin{figure}[t]
  \begin{center}
    \includegraphics[width=0.45\textwidth]{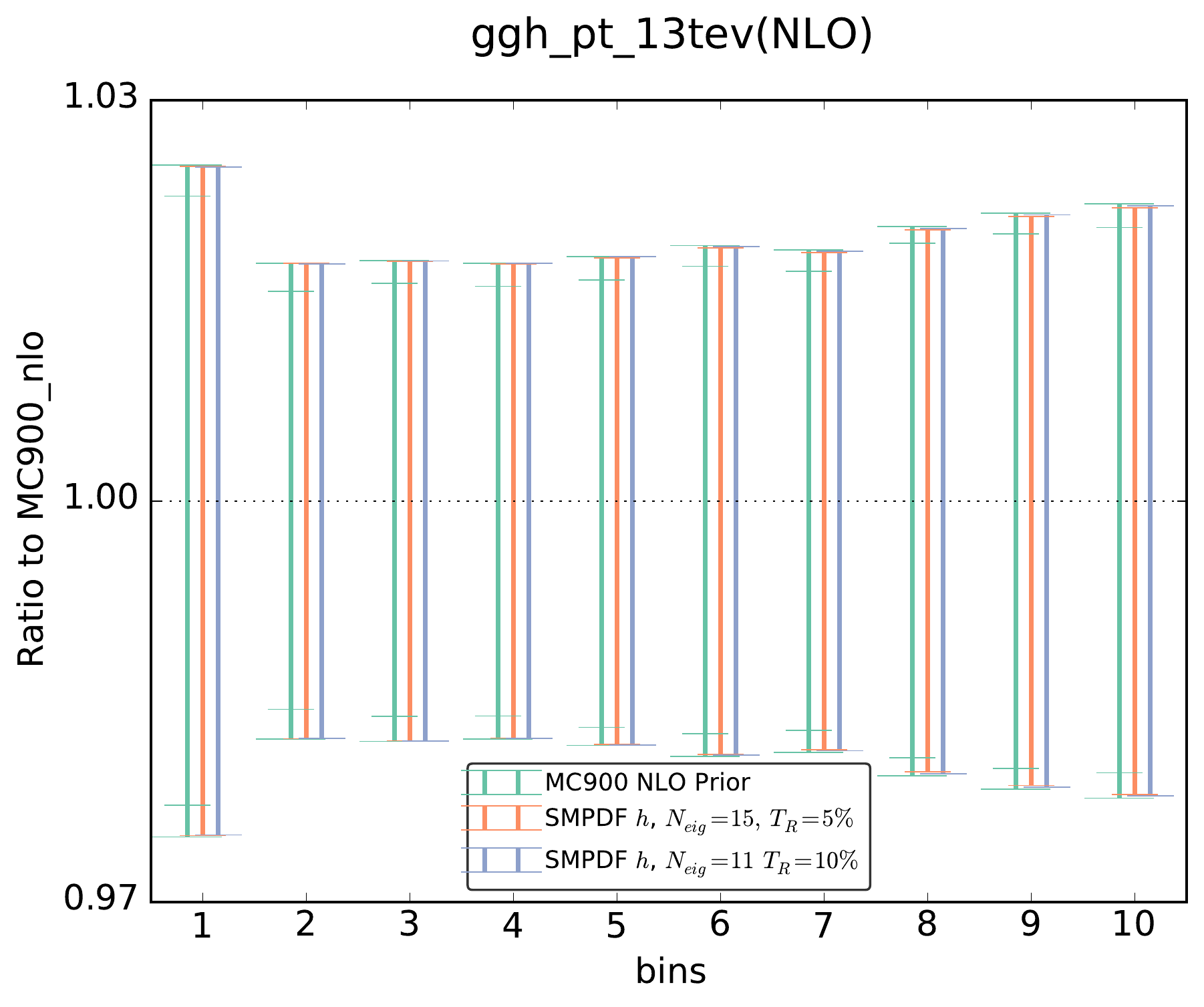}
    \includegraphics[width=0.45\textwidth]{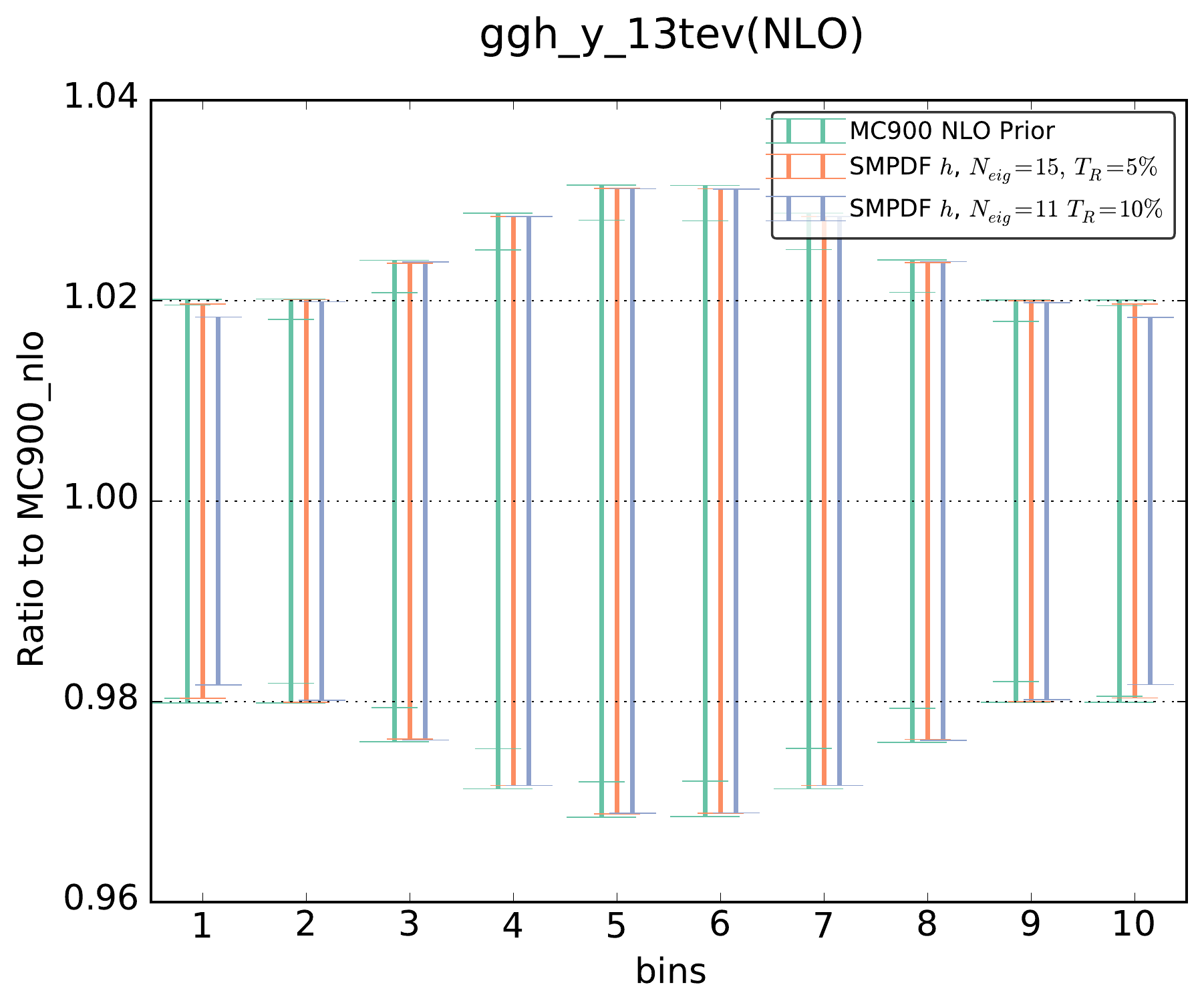}
    \includegraphics[width=0.45\textwidth]{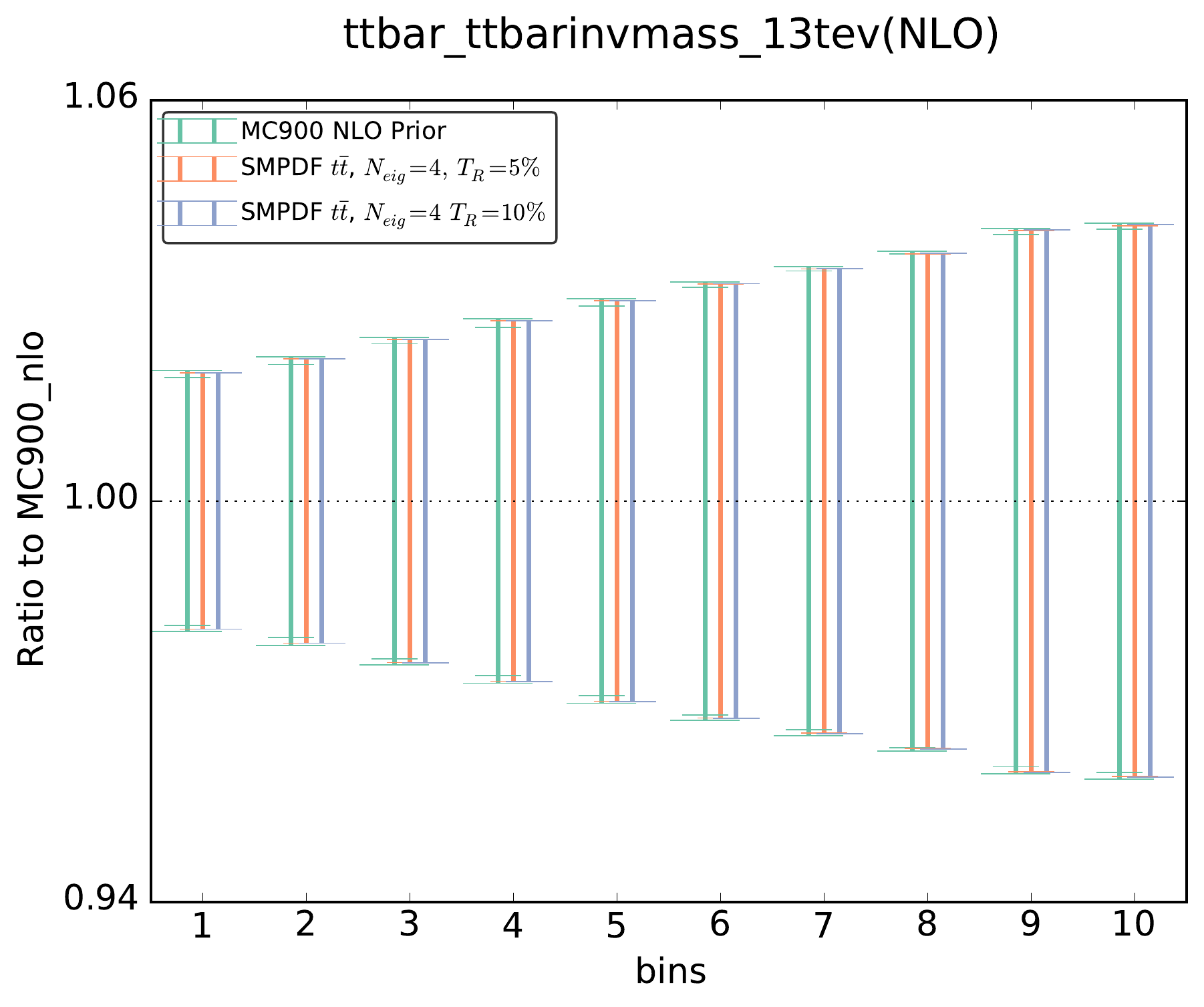}
    \includegraphics[width=0.45\textwidth]{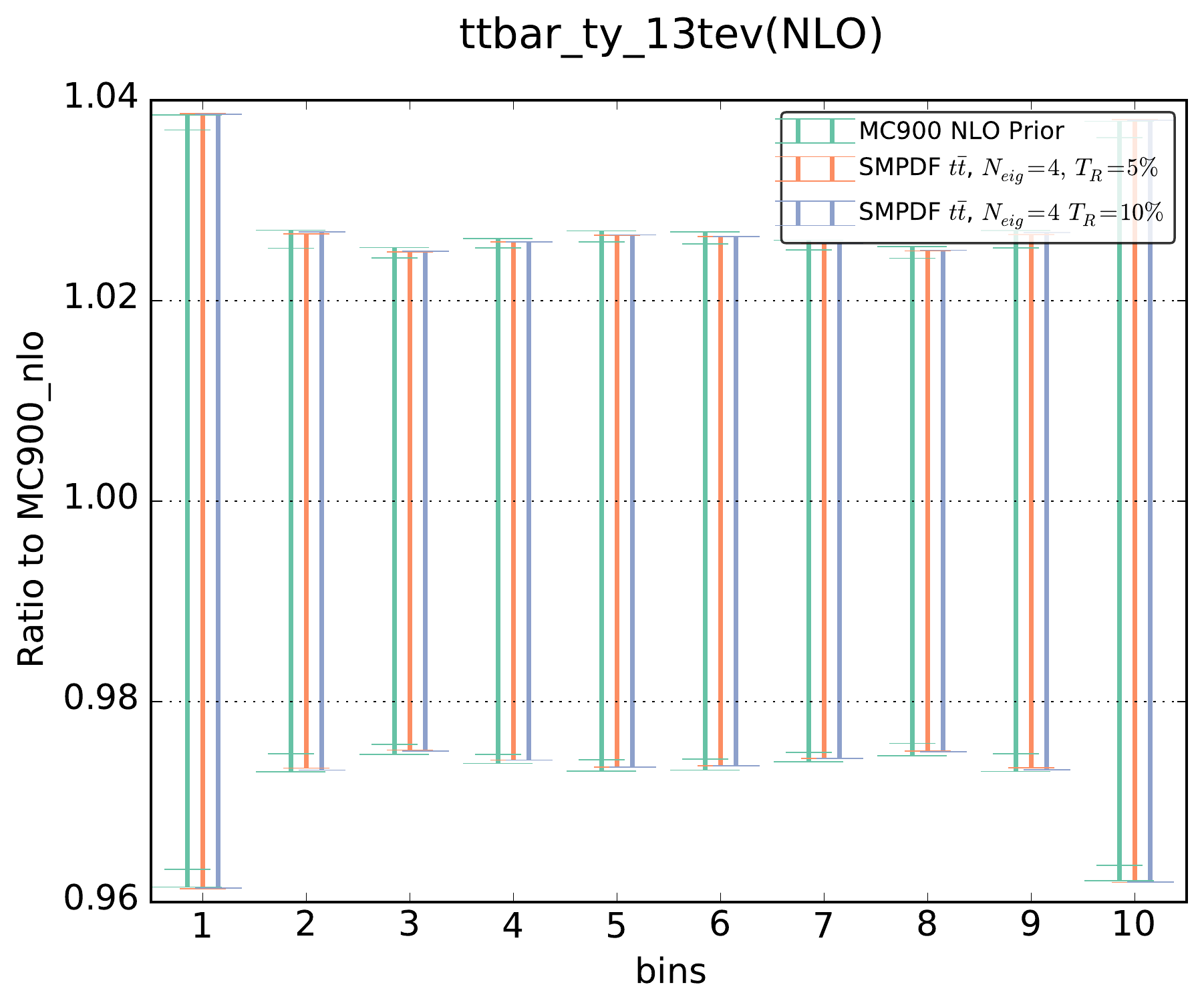}
  \end{center}
  \vspace{-0.3cm}
  \caption{\small Upper plots: comparisons of the
    predictions for the $p_t$
    (left) and rapidity (right) differential
    distributions in Higgs production in gluon fusion
    between the prior MC900 and the corresponding Higgs SM-PDFs
    for two different values of the tolerance $T_R$, 5\% and 10\%.
    Results are shown normalized to the central value of MC900.
    Lower plots: same comparison, now for the $t\bar{t}$ SM-PDFs,
    showing the invariant mass of the $t\bar{t}$ pair $m_{t\bar{t}}$ (left)
    and the top quark rapidity $y^t$ (right).
    See Tables~\ref{tab:processes_H} and~\ref{tab:processes_ttbar}
    for the details of the binning and the kinematical cuts in each case.
  \label{fig:inputpheno}}
\end{figure}

\begin{figure}[t]
  \begin{center}
    \includegraphics[width=0.45\textwidth]{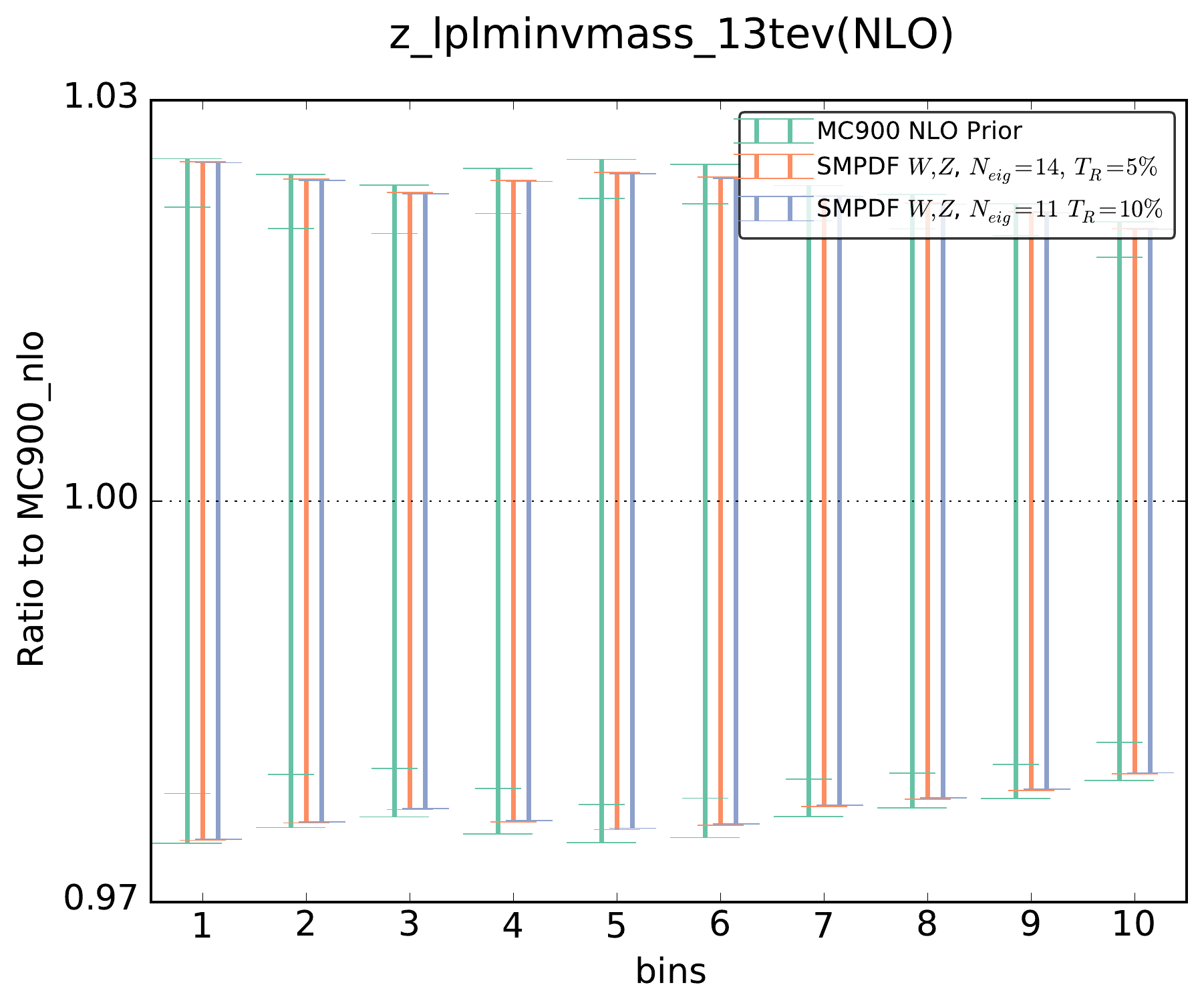}
    \includegraphics[width=0.45\textwidth]{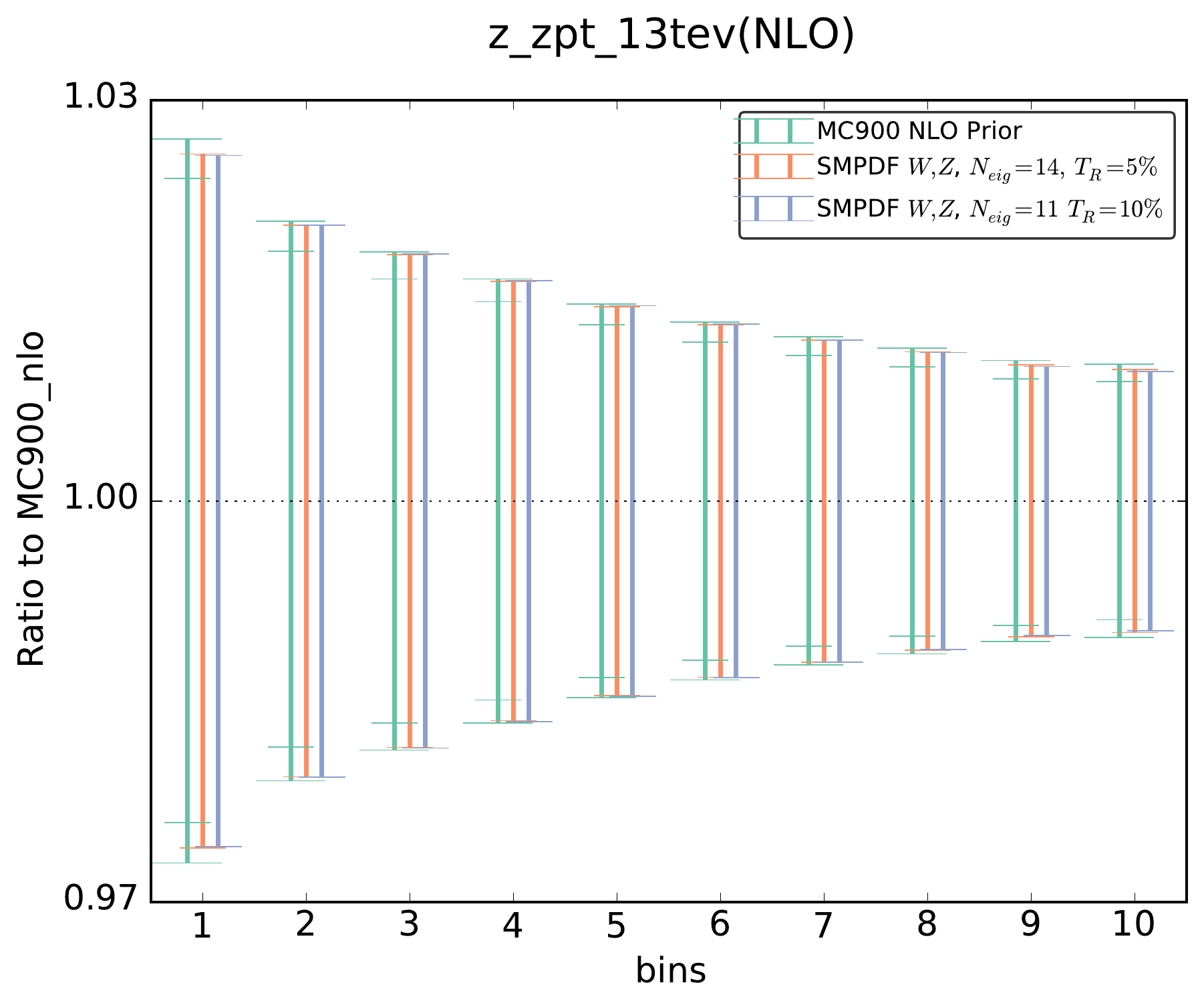}
    \includegraphics[width=0.45\textwidth]{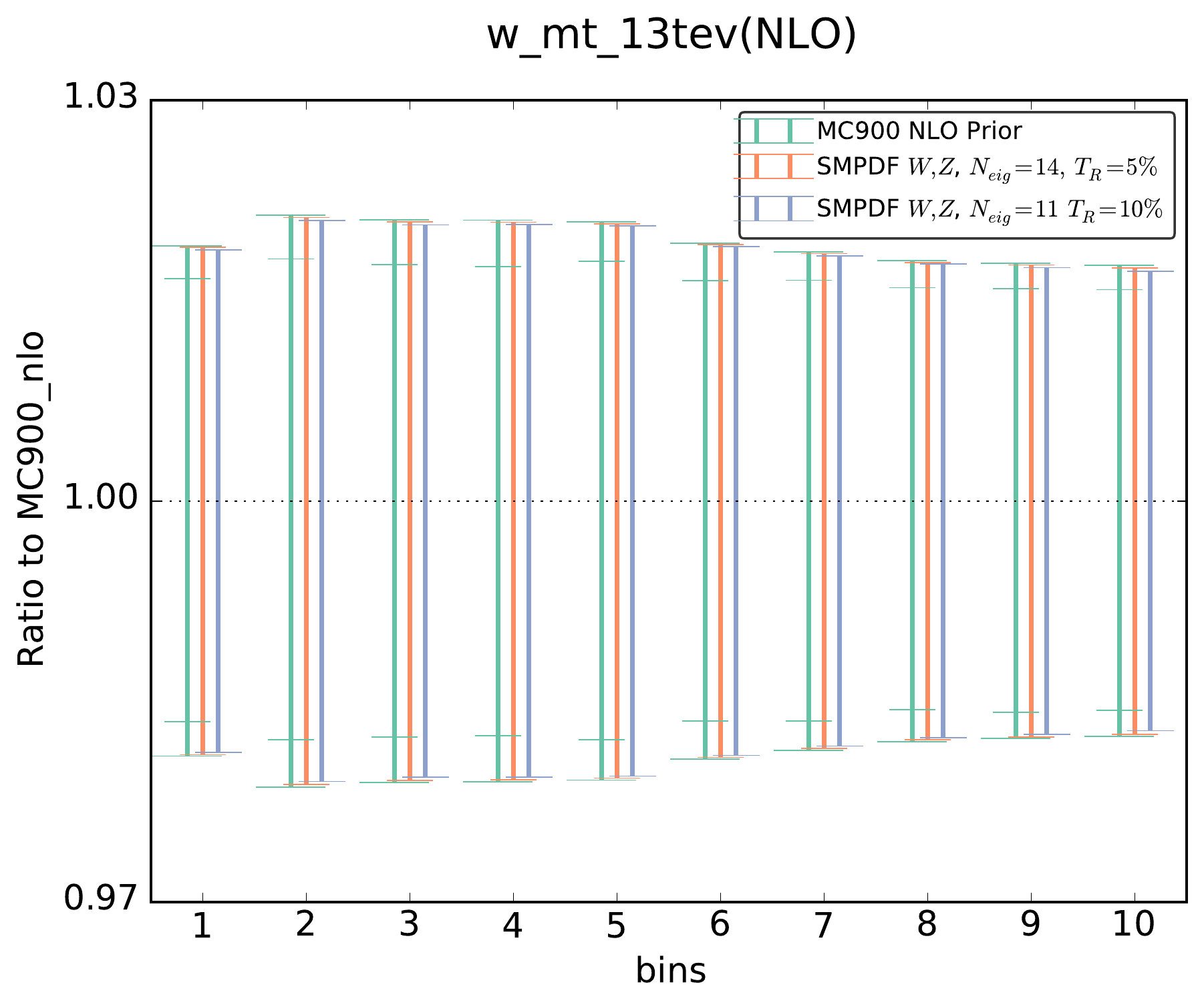}
    \includegraphics[width=0.45\textwidth]{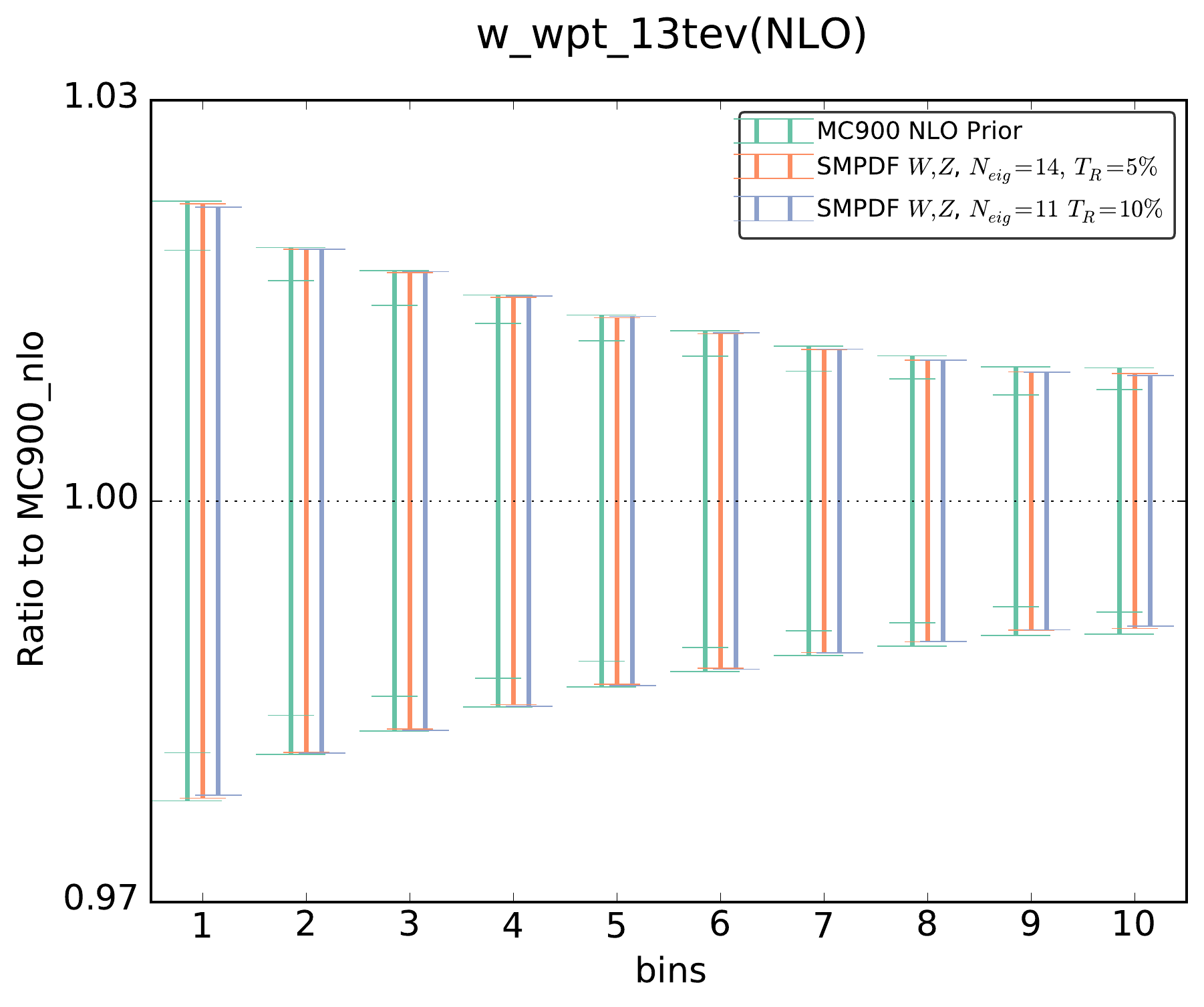}
  \end{center}
  \vspace{-0.3cm}
  \caption{\small Same as Fig.~\ref{fig:inputpheno} for
    representative differential distributions in $W$ and
    $Z$ production, comparing the MC900 prior with the $W,Z$
    SM-PDFs.
  \label{fig:inputpheno2}}
\end{figure}

\begin{figure}[t]
  \begin{center}
    \includegraphics[width=0.45\textwidth]{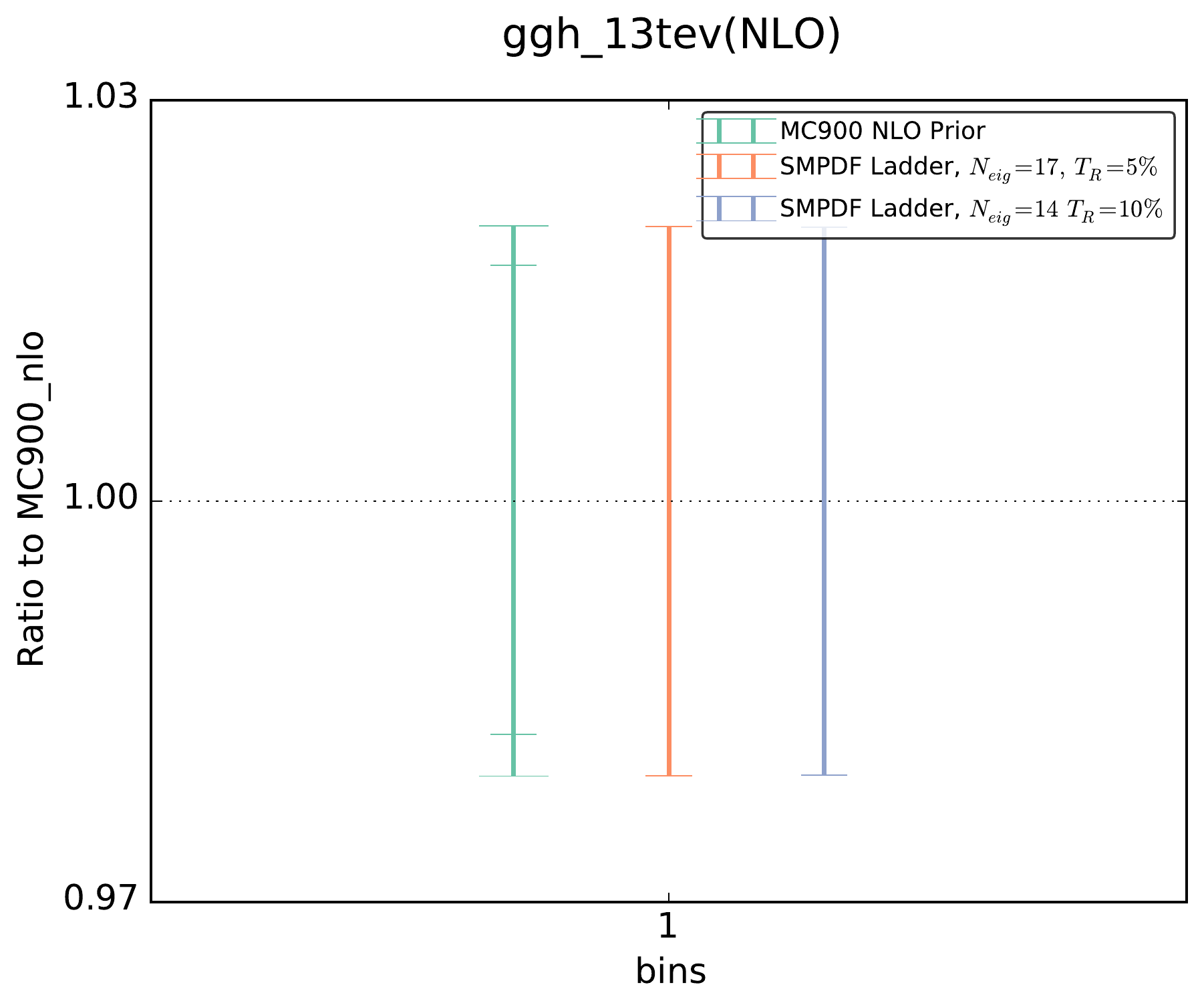}
    \includegraphics[width=0.45\textwidth]{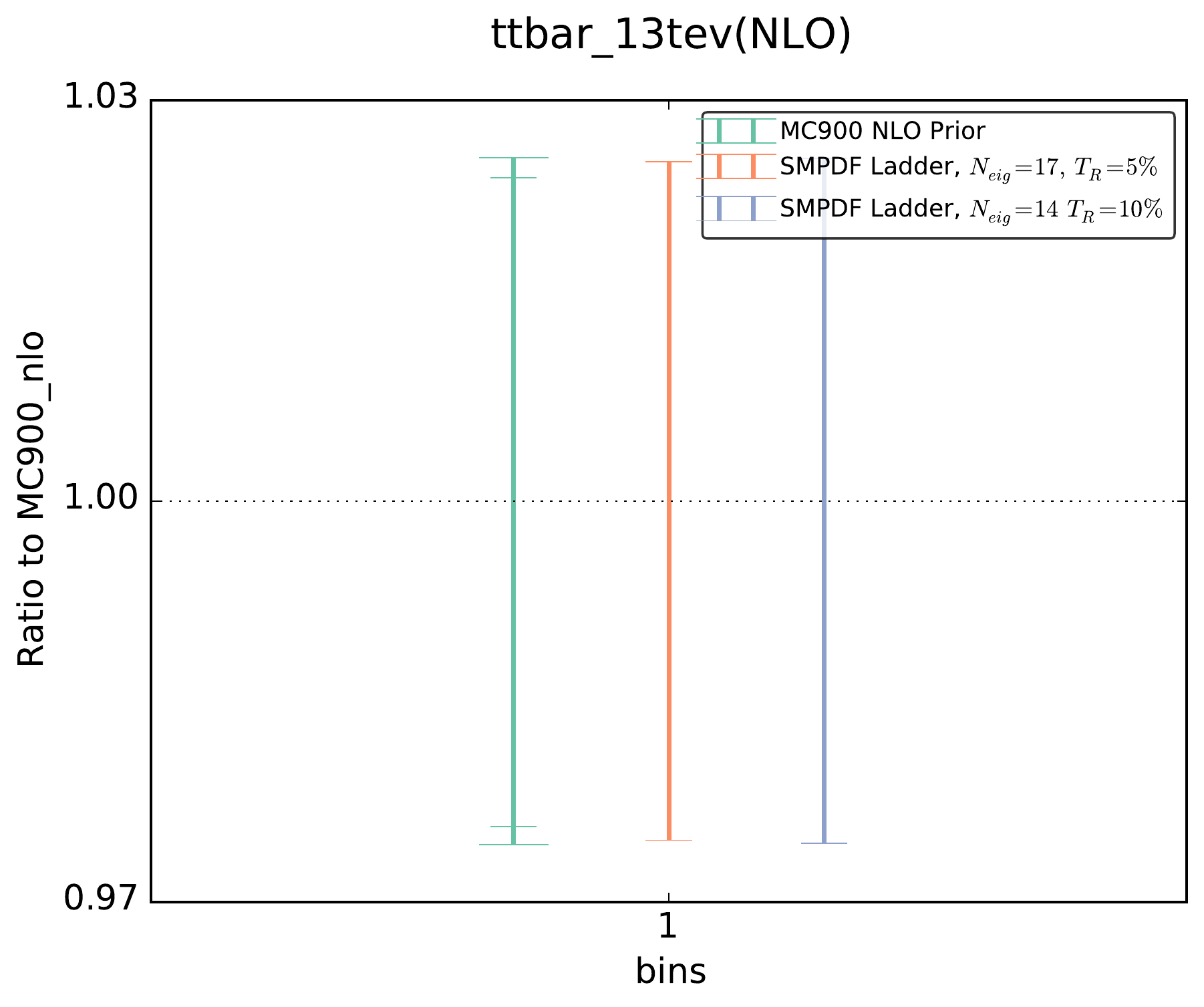}
    \includegraphics[width=0.45\textwidth]{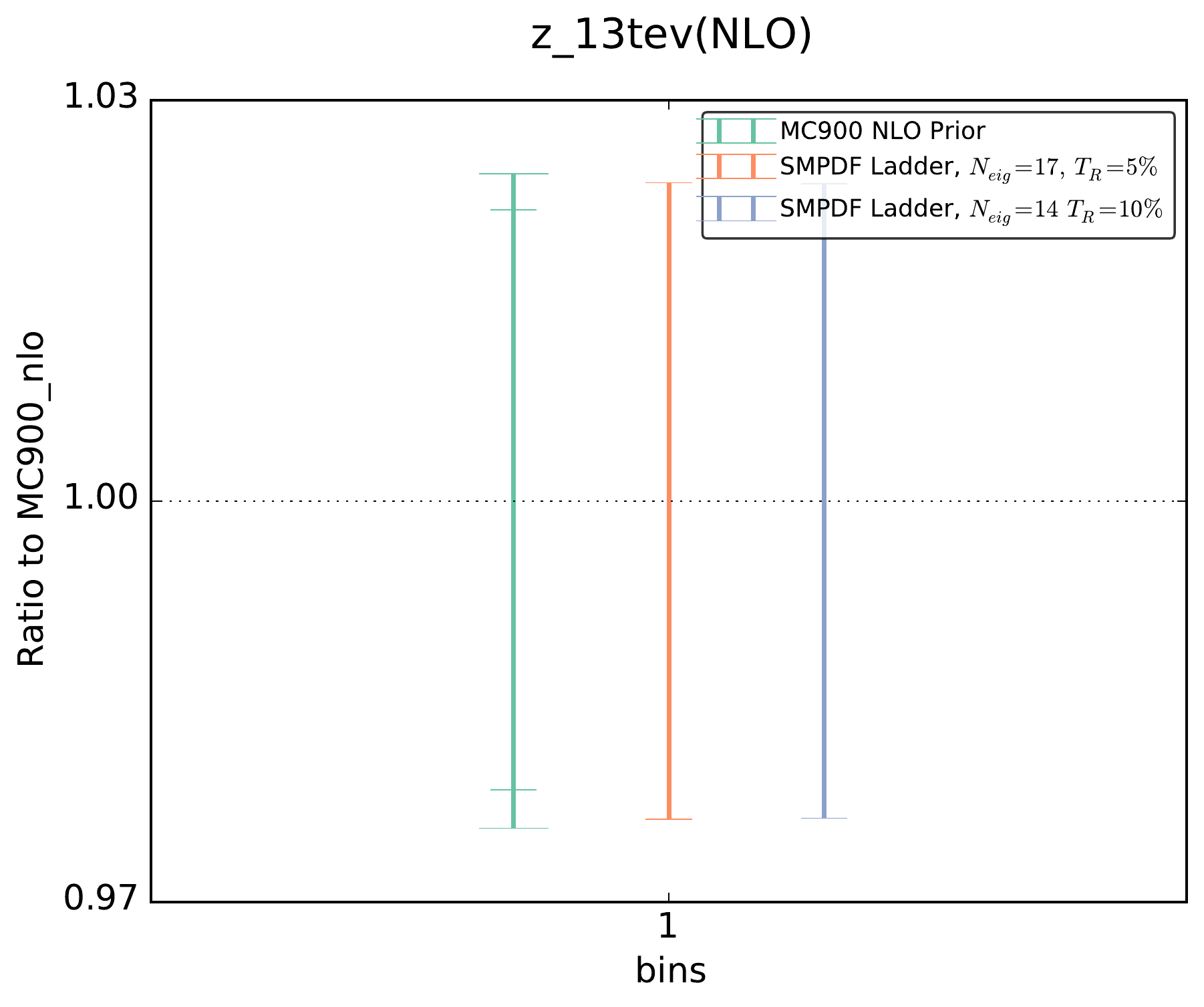}
    \includegraphics[width=0.45\textwidth]{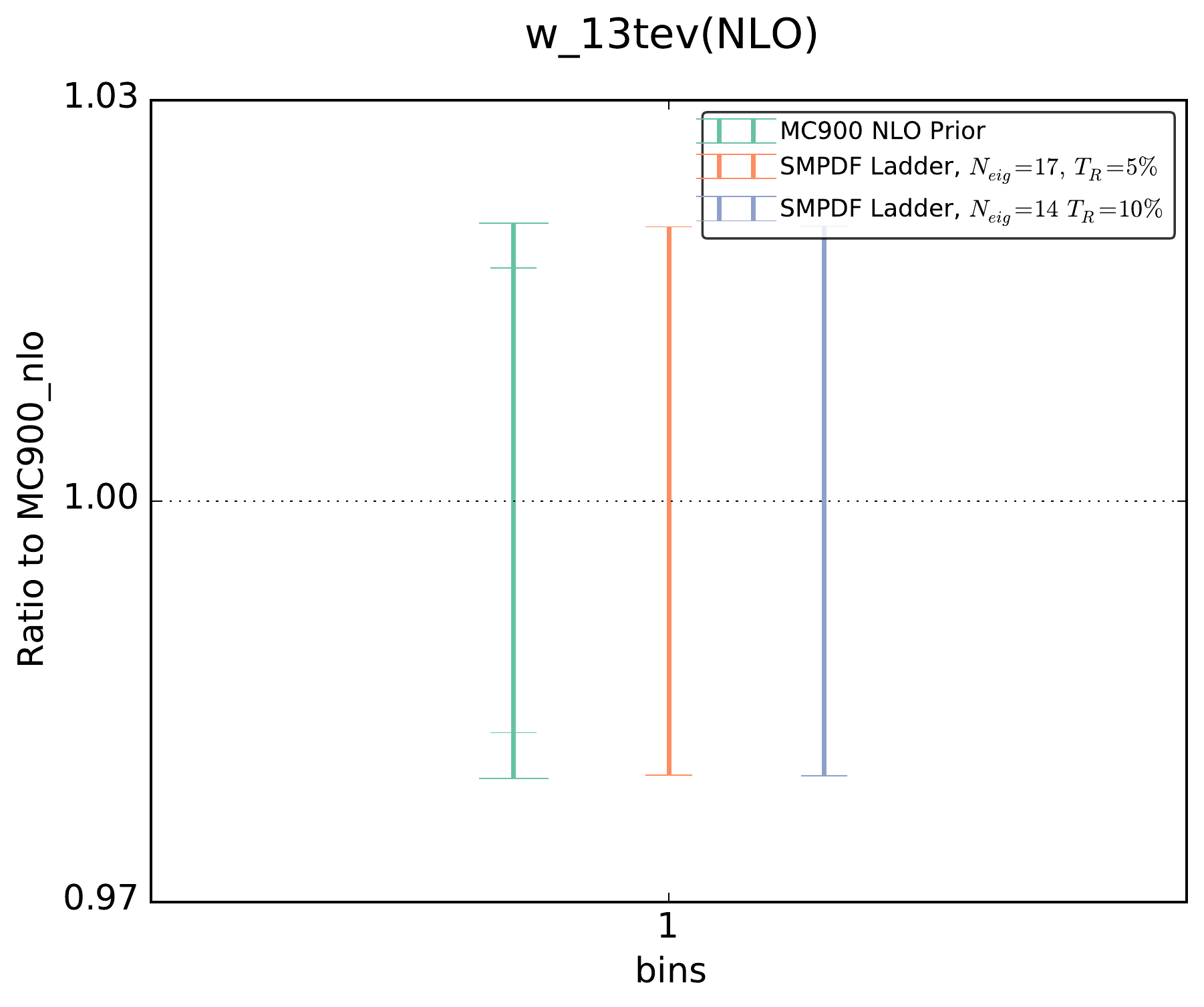}
  \end{center}
  \vspace{-0.3cm}
  \caption{\small Same as Fig.~\ref{fig:inputpheno}
    for the { ``ladder''} SM-PDF, now comparing
    with the total $ggH$, $t\bar{t}$, $Z$ and $W$ inclusive cross-sections.
    \label{fig:ladder}}
  \end{figure}

  In Figs.~\ref{fig:inputpheno}-\ref{fig:ladder}
  we compare various  cross-sections and differential
  distributions
  computed with the MC900 prior PDF set
  and  with the corresponding SM-PDFs for some of the cases
  discussed above, normalized to the
  central value of the prior.
  In the upper plots of Fig.~\ref{fig:inputpheno}, we show
  the Higgs $p_T$ and $y$ distributions in gluon fusion production,
  comparing with the Higgs SM-PDF.
  In the lower plots of Fig.~\ref{fig:inputpheno},
  we show the top quark pair invariant mass $m_{t\bar{t}}$
  and top rapidity $y^t$ distributions, comparing with the
  $t\bar{t}$ SM-PDF.
  In Fig.~\ref{fig:inputpheno2} we compare various differential
  distributions in weak gauge boson production with the
  $W,Z$ SM-PDFs, and finally in
  Fig.~\ref{fig:ladder} we compare the ``ladder'' SM-PDFs
  with various total inclusive cross-sections

  In these comparisons, results are shown for two values of the
  tolerance $T_R=5\%$ and $T_R=10\%$.
PDF uncertainties are shown as  one-sigma
confidence intervals;  for the MC900 prior,
the central 68\% confidence intervals are also
shown (inner ticks).
In all cases we observe excellent agreement between the prior
and the corresponding SM-PDF sets, which provides a further
validation of the reliability of the method.

\begin{figure}[t]
  \begin{center}
\includegraphics[width=0.49\textwidth]{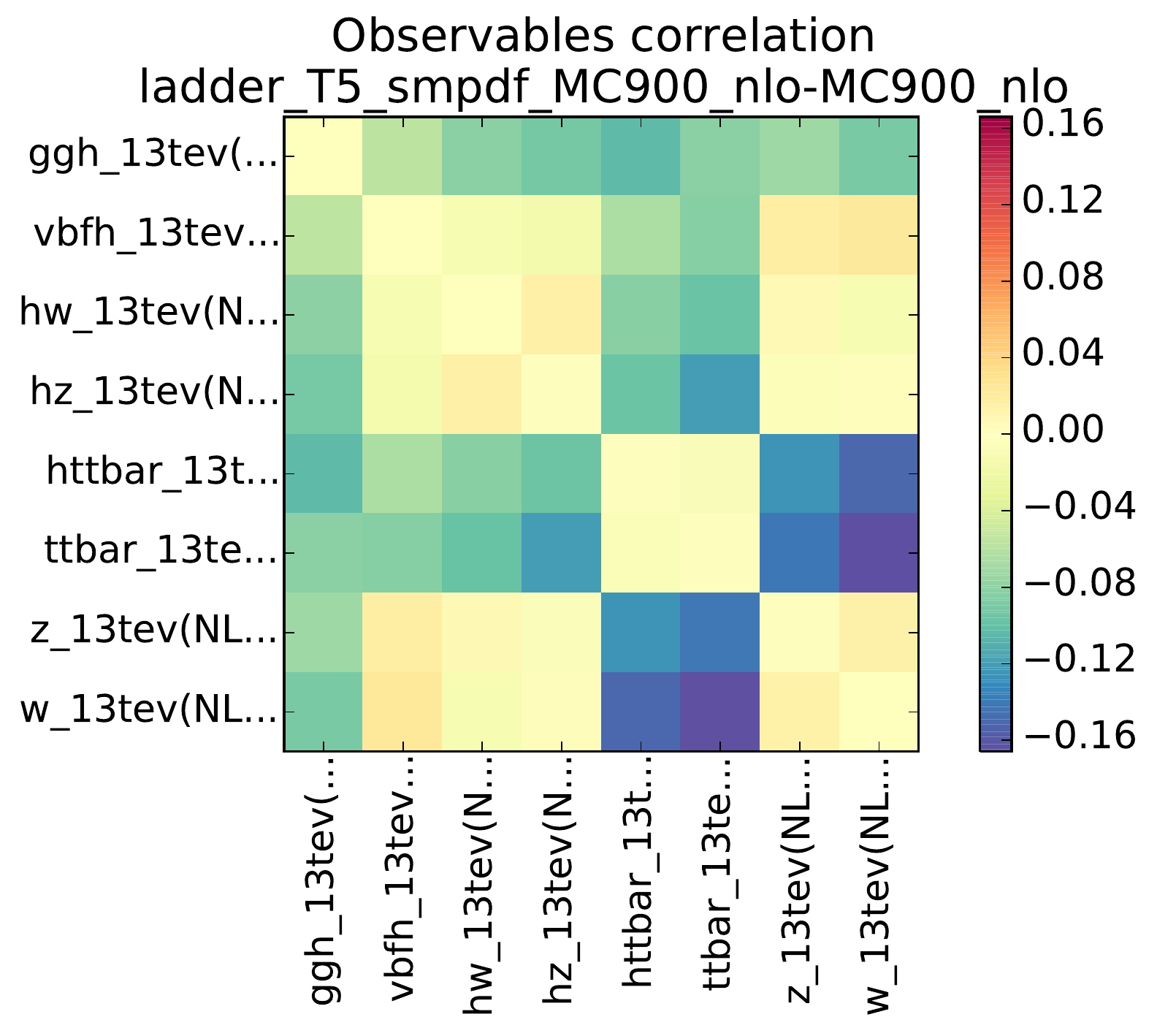}
\end{center}
  \vspace{-0.3cm} \caption{\small Differences in the correlation
    coefficients between the MC900 prior
    and the ``ladder'' SM-PDFs with
    $T_R=5\%$, computed for all the inclusive cross-sections
    that enter the construction of the latter.\label{fig:obscorrs}}
\end{figure}

   We have also verified that SM-PDFs reproduce well PDF correlations,
  even though the tolerance criterion Eq.~(\ref{eq:tolerance}) 
is only imposed on diagonal PDF uncertainties.
  The PDF-induced correlation between two cross-sections computed
  using a Monte Carlo PDF set is given by
  \begin{equation}
    \label{eq:corr_sigma_1}
   \rho(\sigma_i,\sigma_j) =
   \frac{\left<\sigma_1^{(k)}\sigma_2^{(k)}\right>_{\rm rep} -
   \left<\sigma_1^{(k)}\right>_{\rm rep}\left<\sigma_2^{(k)}\right>_{\rm rep}}{s_{\sigma_1}s_{\sigma_2}}\, ,
\end{equation}
while for a Hessian set it is
\begin{equation}
  \label{eq:corr_sigma_2}
   \rho(\sigma_i,\sigma_j) =
   \frac{ \sum_{k=1}^{N_{\rm eig}}\lp \widetilde{\sigma}^{(k)}_i-
     \sigma^{(0)}_i \rp \lp \widetilde{\sigma}^{(k)}_j-
     \sigma^{(0)}_j \rp }{\widetilde{s}_{\sigma_1}\widetilde{s}_{\sigma_2}}\, .
\end{equation}
In Fig.~\ref{fig:obscorrs} we show the
difference between the correlations determined using the 
MC900 prior (from Eq.~(\ref{eq:corr_sigma_1}))
    and the ``ladder'' SM-PDF set (from Eq.~(\ref{eq:corr_sigma_2})), with  
    $T_R=5$\%, for  all the total inclusive cross-sections used as
    input to the
    ``ladder'' SM-PDF set.
We find that the deviation in correlation is 
at the few percent level or better for most cases, and  anyway never
worse than 20\%.

\begin{figure}[t]
  \begin{center}
    \includegraphics[width=0.43\textwidth]{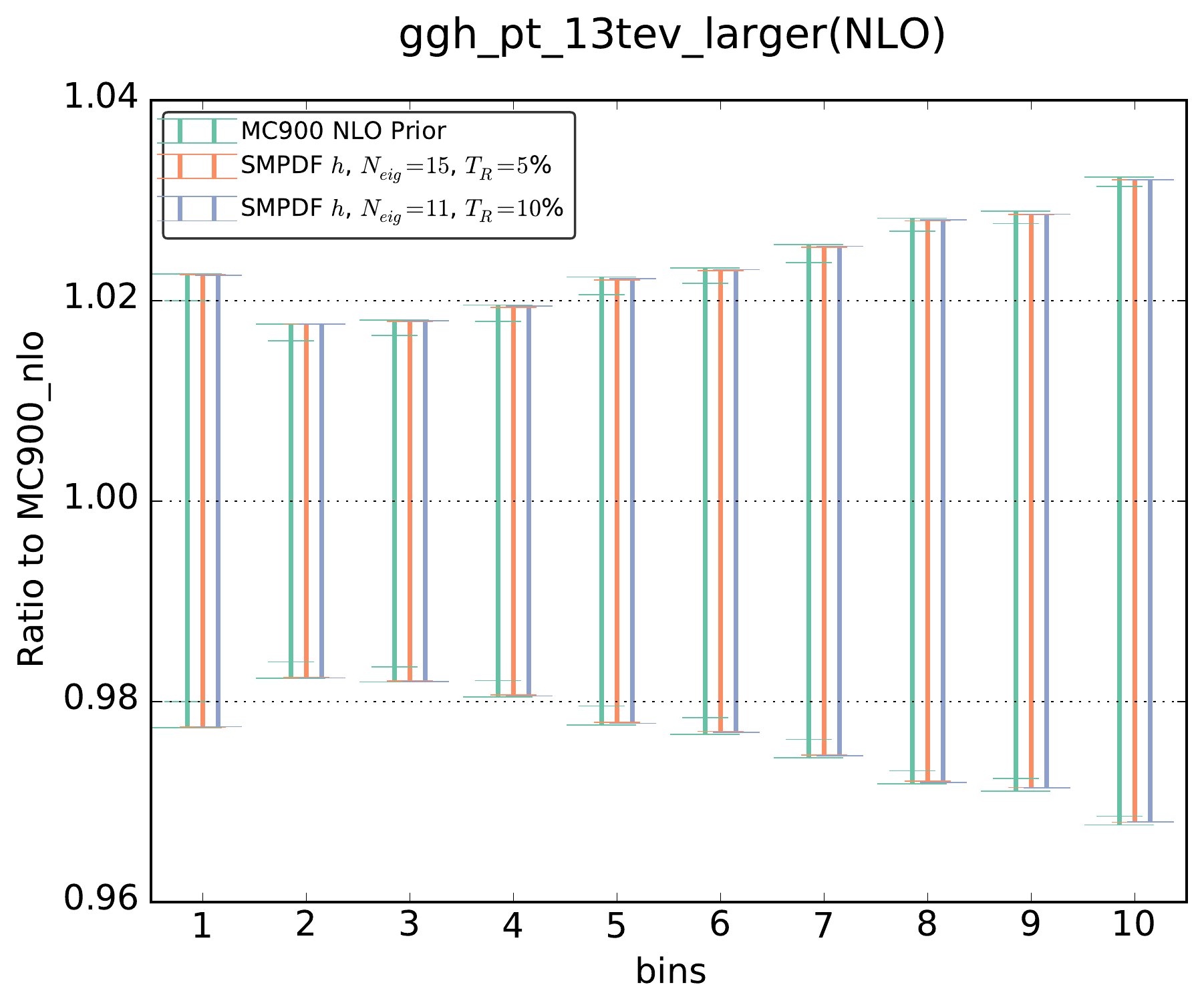}
    \includegraphics[width=0.43\textwidth]{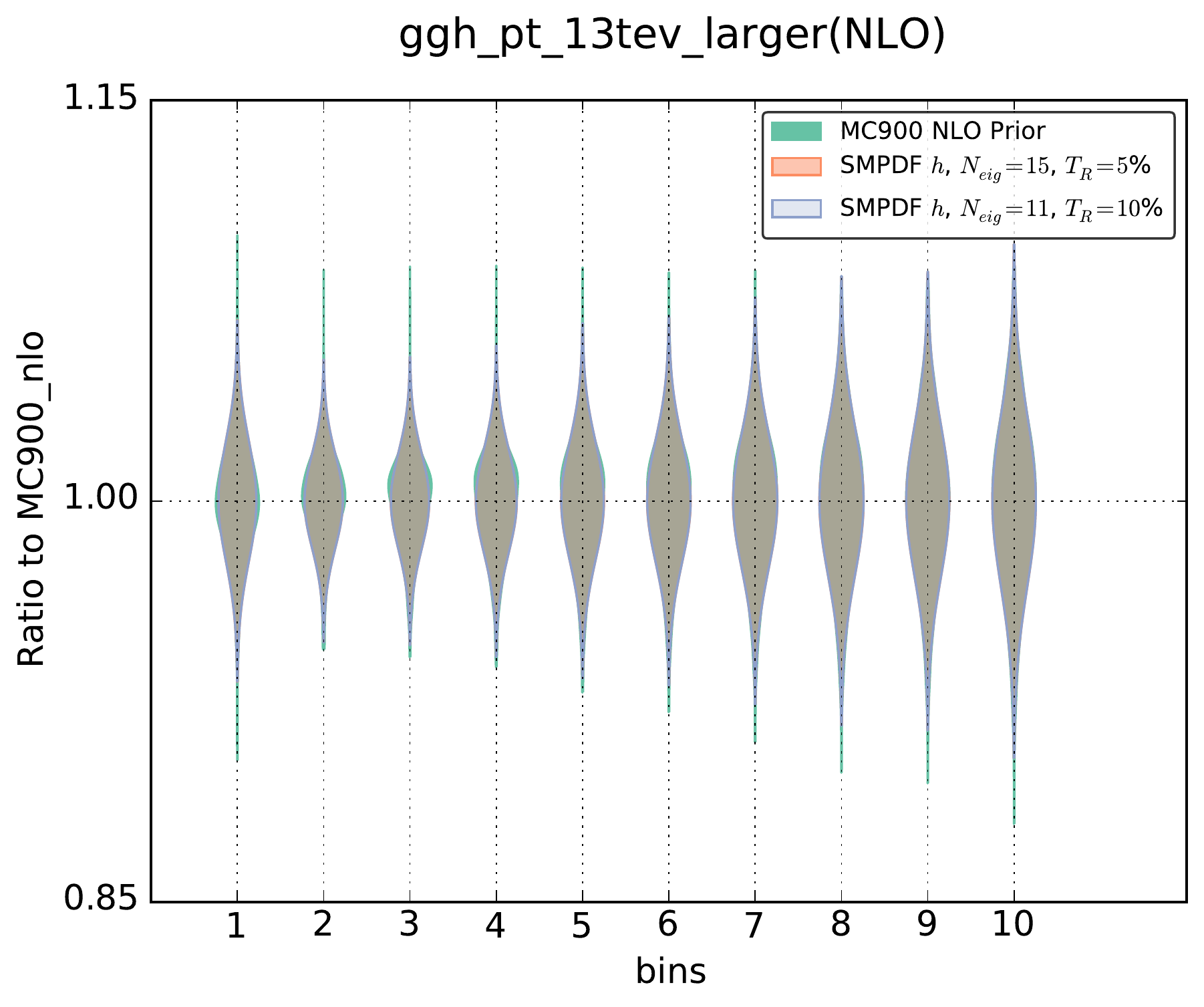}
    \includegraphics[width=0.43\textwidth]{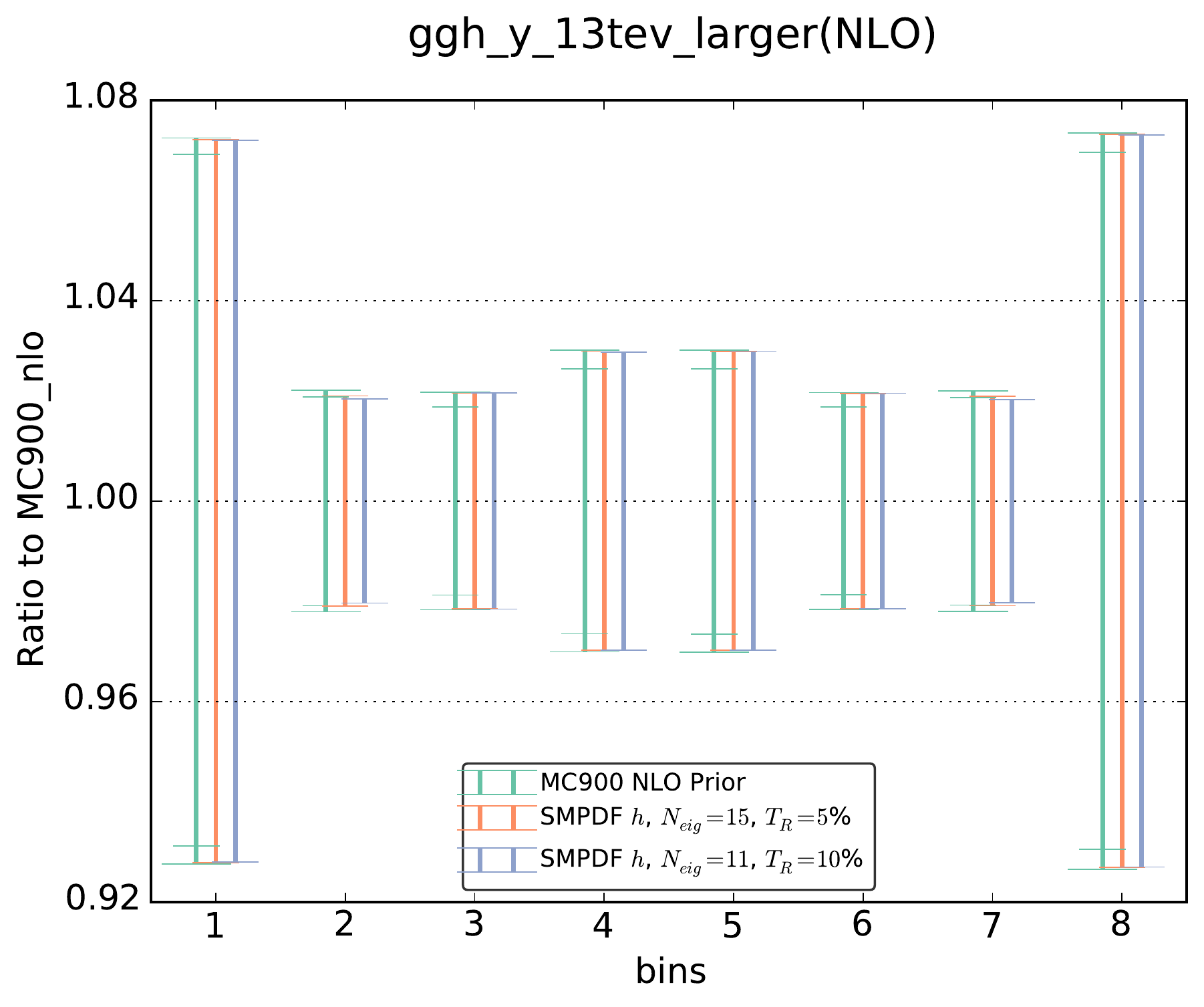}
    \includegraphics[width=0.43\textwidth]{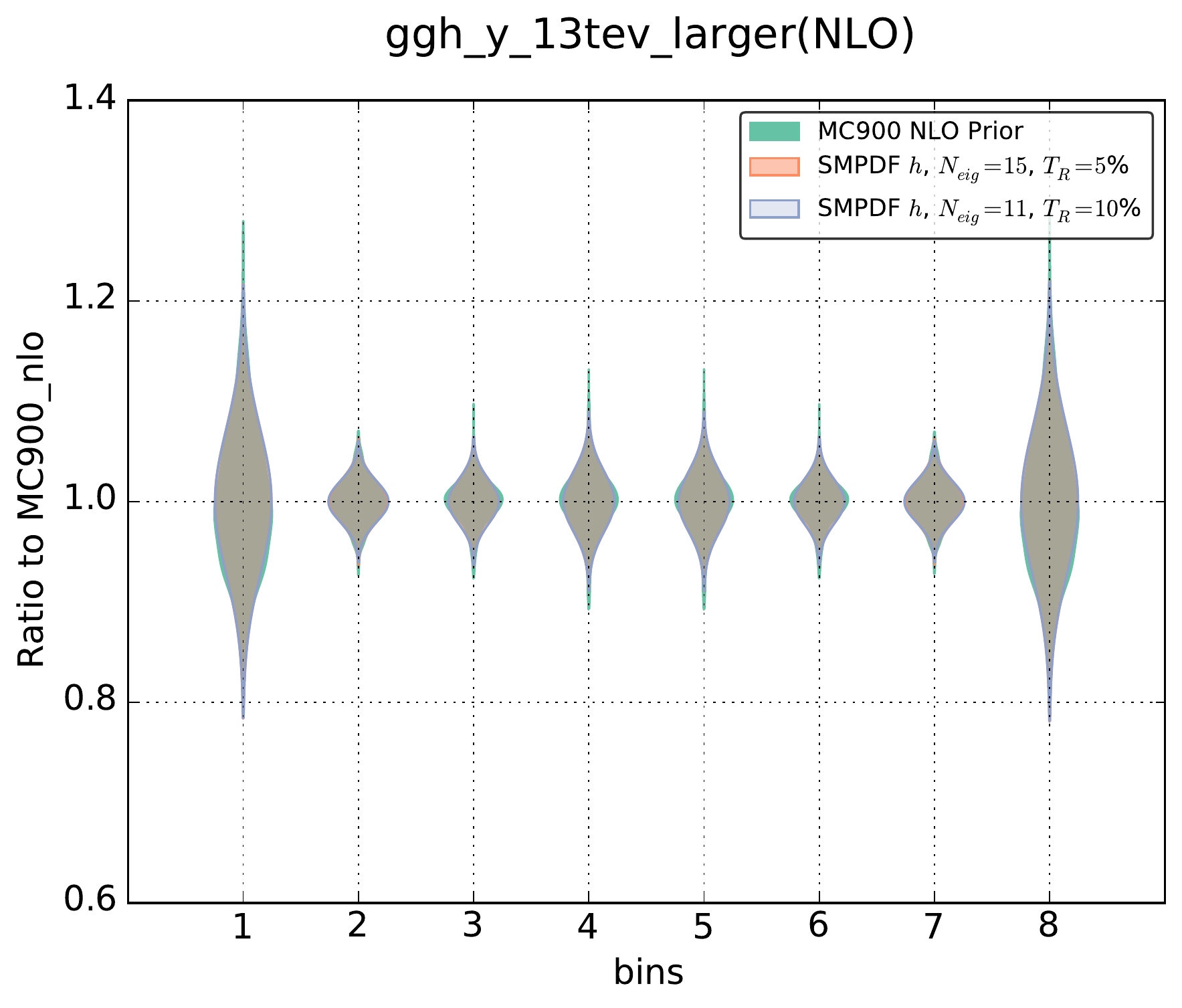}
  \end{center}
  \vspace{-0.8cm}
  \caption{\small The $p_t$ and rapidity distributions
    for Higgs production in gluon fusion, computed with the
    MC900 prior and with the Higgs SM-PDFs, for two values of the tolerance
    $T_R$, this time
    in a kinematic range that doubles that of the input
    processes in Table~\ref{tab:processes_H} (see text).
    In the left plot we show the standard deviation in each bin, while
    in the right plot we show the full probability distributions per bin,
    reconstructed using the Kernel Density Estimate (KDE)
    method.
\label{fig:stress}}
  \end{figure}

An additional validation test can be performed by comparing
the predictions for a given SM-PDF outside the kinematic
range of the input processes.
To illustrate this point, in Fig.~\ref{fig:stress} we compare the $p_t$ and
rapidity distributions in Higgs production via gluon fusion
using the Higgs SM-PDF (which uses as input the processes
in Table~\ref{tab:processes_H}) but
now with an extended kinematical range:
the rapidity distribution now includes $y \in \lc -5,5 \rc$,
rather than  the range $y \in \lc -2.5,2.5 \rc$ used as input,
and the $p_t$ distribution covers now $p_t \in [0,400]$ GeV
as compared to the original input $p_t \in [0,200]$ GeV.
In both cases, we show both the standard deviation (left) and the full
probability distribution obtained with the prior and the two
compressed sets with $T_R=5\%$ and $T_R=10\%$; the smoothened
probability distributions are obtained using the 
using the Kernel Density Estimation (KDE) method discussed
Ref.~\cite{Carrazza:2015hva}.
The good agreement seen in all cases demonstrates the robustness of
the SM-PDF method: namely, SM-PDF sets are stable upon variations of
kinematic cuts and binning of the input cross-sections.

\begin{figure}[t]
    \begin{center}
    \includegraphics[width=0.45\textwidth]{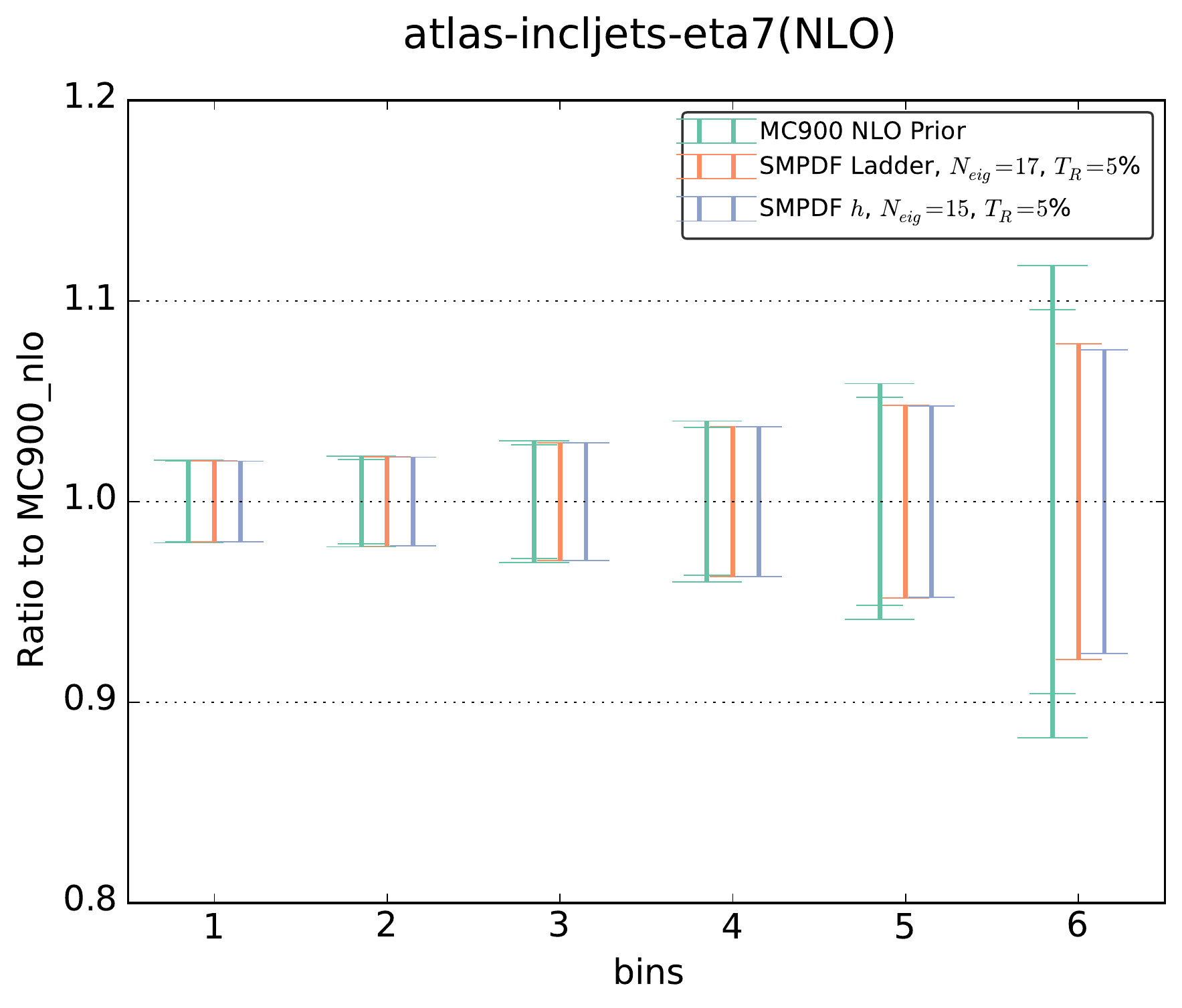}
    \includegraphics[width=0.45\textwidth]{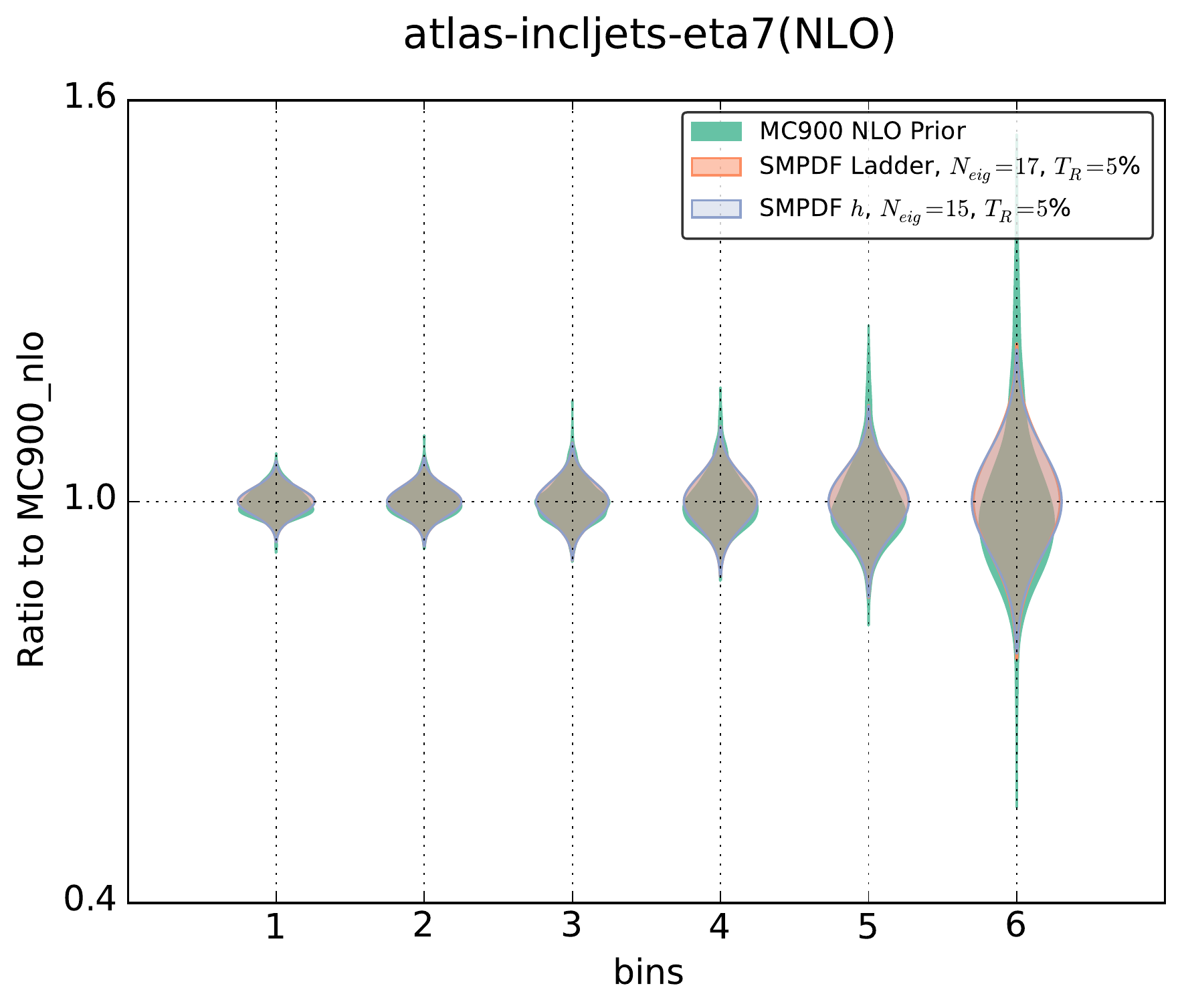}
  \end{center}
  \vspace{-0.8cm}
  \caption{\small Same as Fig.~\ref{fig:stress}, but now for the
ATLAS inclusive jet $p_T$ distribution in
    the  forward region, and using the ``ladder'' SM-PDF set.
    \label{fig:badpred}}
  \end{figure}

While the SM-PDFs are stable upon extrapolation, they will not provide
accurate predictions when used for processes dominated by PDFs in an
altogether different kinematic range.
To illustrate this point,
in Fig.~\ref{fig:badpred} we show
predictions for inclusive jet distributions obtained using the Higgs
and ladder SM-PDF sets, compared to the result obtained using the
MC900 prior.
Specifically, we 
show the 
$p_t^{\rm jet}$ distributions in the
most
forward rapidity bin ($3.6 \le |y_{\rm jet}|<4.4$) of the ATLAS
2010 inclusive jet
measurement~\cite{Aad:2011fc}; bins are ordered in increasing
$p_T$.
Clearly, the agreement deteriorates at large $p_T$, where
results depend on the large-$x$ quarks and gluon, which are
weakly correlated to the processes included in the construction of the
both the Higgs
and ``ladder'' SM-PDF sets. This also suggests that good agreement,
with a marginally larger number of eigenvectors, could be likely
obtained by just widening the range of some of the inputs to the
``ladder'', such as, for instance, including the Higgs
  transverse-momentum distribution up higher values of $p_t$.
In fact, we have explicitly checked~\cite{leshouches} 
that the ``ladder'' PDF set
provides comparable accuracy to the PDF4LHC15 30 eigenvector set when
used for the determination of 
all the hadronic observables included in the NNPDF3.0 PDF
determination~\cite{Ball:2014uwa}, despite having almost half the
number of eigenvectors. 

\clearpage

\section{A posteriori combination of SM-PDFs}
\label{sec:combination}

So far, we considered the construction of a PDF set tailored to  a given list
of input cross-sections.
However, one may also
encounter the situation in which two SM-PDF sets constructed using
different processes as input are already available, and wishes to use them
simultaneously, without having to
produce a new  dedicated SM-PDF set using as input the two processes at the same time.
A typical application is a  computation in which one of these
processes is the signal and the other to a background.
The SM-PDF methodology also
allows to deal with this situation: we first discuss how this is done,
and then we present an example of application.

\subsection{General method}
\label{sec:combmeth}

In Sect.~\ref{sec:methodology} we have shown how, starting
from a Monte Carlo PDF prior, $X_{lk}$, Eq.~(\ref{eq:Xmat}),
we can construct a specialized minimal Hessian representation,
$\widetilde{X}_{lk}$, Eq.~(\ref{eq:newX}), in terms of
a reasonably small number of eigenvectors.
The
result of the  SM-PDF
algorithm can be expressed as a regular Hessian PDF set, with the error
parameters given by Eq.~(\ref{eq:newhessian}).
Alternatively, one can 
directly use the final matrix of Hessian coefficients $P$ to express
the cross-sections computed with each of the replicas of the prior
set, Eq.~(\ref{eq:obsdiffs}), as linear combinations of cross-sections
computed with the final eigenvector sets,
Eq.~(\ref{eq:diffpredictediter}). The two results are equivalent  by 
linear error propagation.

However, we can also read  Eq.~(\ref{eq:diffpredictediter}) in reverse: if
we define 
\begin{equation}	
   \label{eq:combine}
    d_k^{\rm MC}(\sigma_i) = \sqrt{N_{\rm rep}-1} \sum_{j=1}^{N_{\rm eig}}P_{kj}{d^P}_j(\sigma_i)  \, , \qquad k=1,\ldots,N_{\rm rep} \, , \qquad  i=1,\ldots,N_{\sigma} \, .
\end{equation}
we can view the set of $N_{\rm rep}$  differences
$d_k^{\rm MC}(\sigma_i)$ (for each of the $N_{\sigma}$ observables $\sigma_i$)
as a Monte Carlo set of cross-sections, containing the same
information as the reduced SM-PDF set.
In other words, the  $N_{\rm rep}$ values
\begin{equation}\label{eq:pseudomc}
\sigma^{(k)}_i
=\sqrt{N_{\rm rep}-1}\sum_{j=1}^{N_{\rm eig}}P_{kj}{d^P}_j(\sigma_{i})+\sigma_{i}^{(0)}
\, , \qquad k=1,\ldots,N_{\rm rep} 
\, , \qquad  i=1,\ldots,N_{\sigma} \, ,
\end{equation}
of the observable $\sigma_i$ can be viewed as ``pseudo-Monte Carlo''
replicas, to be used to compute uncertainties and correlations using
the standard Monte Carlo procedure.

If two sets of SM-PDFs corresponding to different processes are
available, we can then combine the information contained in them  
by first turning the
predictions obtained from them into
replicas using Eq.~(\ref{eq:pseudomc}), and then viewing the set of
Monte Carlo replica predictions obtained in each case as our best approximation
to the Monte Carlo set of predictions for that process obtained with the
original PDF replica set.  These sets of prediction replicas can
then be used in order to compute any quantity which depends on both processes
 using standard Monte Carlo methodology, by just making sure that
each process is computed using its corresponding replicas.

\subsection{Validation}

We illustrate and validate the methodology presented in
Section~\ref{sec:combmeth} with an example.
We use as input prior the NNPDF3.0 NLO set with
$N_{\rm rep}=1000$ replicas and then
generate two SM-PDFs for a fixed choice of the
tolerance $T_R=5\%$.
The first SM-PDF takes as input the $t\bar{t}$ processes from
Table~\ref{tab:processes_ttbar}, while the second is constructed
from the $W,Z$ processes of Table~\ref{tab:processes_WZ}.

We now use these two SM-PDF sets to calculate the PDF
uncertainties on the $t\bar{t}$ and the $W$ total inclusive cross sections.
This can be done both with the original
representation, Eq.~(\ref{eq:obsstd}), or with
the new SM-PDF Hessian representation.
As
shown in Table \ref{tab:neig}, we find $N_{\rm eig}=5$ for the
$t\bar{t}$ SM-PDF and $N_{\rm eig}=13$ for the $W,Z$ SM-PDF.
We obtain the following results for the total cross-sections:
for the $t\bar{t}$ cross-section with $t\bar{t}$ SM-PDFs 
\begin{equation}
	\sigma_{t\bar{t}\ ({\rm prior})} = 671.12\pm12.0\ {\rm pb} \, ,
\end{equation}
\begin{equation}
	\sigma_{t\bar{t}\ ({\rm smpdf-tt})} = 671.12\pm11.9\ {\rm pb} \, ,
\end{equation}
and for the $W$ cross section with $W,Z$
SM-PDF
\begin{equation}
	\sigma_{W\ (\rm prior)} = 23867\pm419\ {\rm pb}\,,
\end{equation}
\begin{equation}
	\sigma_{W\ (\rm smpdf-wz)} = 23867\pm417\ {\rm pb} \, .
\end{equation}

Now  suppose that we want to  compute a quantity which depends both on
$t\bar{t}$ and $W$ cross-sections, such as the ratio between the two,
$\sigma_{t\bar{t}}/\sigma_W$.
In the computation of the PDF uncertainty  on this ratio, it is essential
to properly account for the cross-correlations between the two processes.
This can be achieved by recasting
 the results of the two different SM-PDFs into corresponding
Monte Carlo sets of predictions through Eq.~(\ref{eq:pseudomc}).

Namely, the PDF uncertainty on the
cross-section ratio is given by
\begin{equation}
	\label{eq:ratiomc}
    s_{\frac{\sigma_{t\bar{t}}}{\sigma_{W}}}=
        \frac{1}{N_{\rm rep}-1}\left(\sum_{k=1}^{N_{\rm rep}}
        \left(\frac{\sigma^{(k)}_{t\bar{t}}}{\sigma^{(k)}_{W}}-
        \left\langle\frac{\sigma^{(k)}_{t\bar{t}}}{\sigma^{(k)}_{W}}
		\right\rangle_{\rm rep}
        \right)^{2}\right)^{\frac{1}{2}}, 
\end{equation}
where $\sigma^{(k)}_{t\bar{t}}$ and $\sigma^{(k)}_{W}$ have been
obtained using Eq.~(\ref{eq:pseudomc})  with the $P$ matrix that
corresponds respectively to the $t\bar{t}$ and $W,Z$ SM-PDF sets. 

Using Eq.~(\ref{eq:ratiomc}) we get
\begin{equation}
	\label{eq:ratioresult}
	s_{\frac{\sigma_{t\bar{t}}}{\sigma_{W}}}= 6.66497\times 10^{-4} \,,
\end{equation}
to be  compared to the result obtained  from the NNPDF3.0 prior, using the $N_{\rm rep}=1000$
original replicas,
\begin{equation}
	\label{eq:ratioresultB}
	s_{\frac{\sigma_{t\bar{t}}}{\sigma_{W}}({\rm prior})}= 
	6.66503\times 10^{-4}\ ,
\end{equation}
which is identical for all practical purposes.

It is important to realize  
that while Eq.~(\ref{eq:ratioresultB}) requires the calculation of
$2N_{\rm rep}=2000$ cross-sections,  Eq.~(\ref{eq:ratioresult}) 
only requires the knowledge of the $N_{\rm eig}$ cross-section differences
$\widetilde{d}_j(\sigma_i)$ for the two observables, which is equal to
the sum of the number of eigenvectors in the two sets which are being
combined, in our case, $N_{\rm eig}^{WZ}+N_{\rm eig}^{t\bar t}=18$,
with great computational advantage.

As a further cross-check, we have recomputed the same cross-section ratio by 
using the methodology of Sect.~\ref{sec:methodology}, namely, by
constructing a dedicated SM-PDF set
using
as input the two families of processes, $t\bar{t}$ and $W,Z$, simultaneously.
This new SM-PDF has now 17 eigenvectors for the case of a tolerance
$T_R=5\%$
and leads to
\begin{equation}
	\label{eq:ratioresultC}
	s_{\frac{\sigma_{t\bar{t}}}{\sigma_{W}}({\rm combined})}= 
	6.655\times 10^{-4}.
\end{equation}

This shows that  the
advantage of constructing a dedicated set in comparison to combining
the pre-existing sets is marginal, as the accuracy is the same, and the total number of eigenvectors
$N_{\rm eig}$ 
has only decreased by one unit.

\section{Delivery}
\label{sec:delivery}

Building upon our previous MC2H methodology for the construction of
reduced Hessian representations of PDF
uncertainties~\cite{Carrazza:2015aoa}, we have presented an algorithm
for
 the construction of a minimal Hessian representation of any
given prior PDF set, specialized to reproduce a number
of input cross-sections.
We have shown that the algorithm can be used to
construct specialized minimal PDF sets which reproduce
with percent accuracy the central values and
PDF uncertainties for all input observables
in terms of a substantially smaller number of eigenvectors
as compared to the prior PDF set.
A remarkable advantage of the SM-PDF methodology is that the
complete information contained in the original prior set is kept at all stages
of the procedure. As a consequence, it is possible to add new
processes to any given SM-PDF set with no information loss.
Also, it
is possible to combine {\it a posteriori} SM-PDF sets corresponding to different
processes without any new computation.

The  SM-PDF code is publicly available  from the repository
\begin{center}
\url{https://github.com/scarrazza/smpdf/}
\end{center}
The code is  written in {\tt Python} using the numerical
implementations provided by the {\tt NumPy} package.
Customized interfaces to {\tt APPLgrid} and {\tt LHAPDF6} are also
included.
The package also includes the {\tt APPLgrid} grids for all the processes
listed in Tables~\ref{tab:processes_H} to~\ref{tab:processes_WZ},
and additional processes can be easily generated upon request.

The input of the  SM-PDF code is the prior PDF set
and the list of cross-sections $\{ \sigma_i\}$ to
be reproduced.
The code settings can be
be modified by the user by means of a steering card.
The cross-sections can be provided either by indicating the
name of the {\tt APPLgrid} or by means of a text file (for
predictions computed with external codes).
An example steering card for the code is presented
in Appendix~\ref{sec-appendix}.

The output of the code is then the corresponding SM-PDF set,
directly in the
{\tt LHAPDF6} format, as well as the corresponding
direct and inverse Hessian parameter matrices, $P$ and
$P^t$, respectively as a {\tt CSV} file.
These rotation matrices allow to easily transform
computed cross-sections back and forth from any SM-PDF
representation to the prior representation, as well as transforming
between different SM-PDF representations, as explained
in Sect.~\ref{sec:combination}.

Together with this, a number of additional validation
features are included in the  SM-PDF package.
In particular, comparisons at the level of the
input cross-sections as those presented in
Figs.~\ref{fig:corrthreshold},~\ref{fig:inputpheno} and~\ref{fig:obscorrs}
can be generated automatically by filling the appropriate
options in the {\tt YAML}
configuration file, without the need
of writing additional code.
The user is encouraged to refer to the documentation for a more extensive
description of the different features available.
In addition, a web interface to similar to that of
{\tt APFEL Web} on-line
PDF plotter~\cite{Bertone:2013vaa,Carrazza:2014gfa} is currently under
consideration.

Finally, the SM-PDFs constructed in Sect.~\ref{sec:validation}
are also available from the same webpage in the {\tt LHAPDF6}
format.
Users can produce the SM-PDFs that more most suitable for
specific applications
by generating the suitable cross-section
theory calculations and then running the  SM-PDF code. However, users
are  encouraged to contact the authors for support if assistance is needed.
Additional SM-PDFs can be  added to this webpage upon request.

\section*{Acknowledgments}

We wish to thank Andr\'e David for helpful ideas an explanation on the
experimental requirements of specialized minimal PDF sets.
We are grateful to Jun Gao, Joey Huston, Pavel Nadolsky, Robert
Thorne and other colleagues of the PDF4LHC Working Group for
illuminating discussions on the topic of PDF reduction
strategies.

S.~C. and S.~F. are supported in part by an Italian PRIN2010 grant and
by a European Investment Bank EIBURS grant.
S.~C. is supported by the HICCUP ERC Consolidator grant (614577).
S.~F. and Z.~K. are
supported by the Executive Research Agency (REA) of the European
Commission under the Grant Agreement PITN-GA-2012-316704 (HiggsTools).
J.~R. is supported by an STFC Rutherford Fellowship
and Grant ST/K005227/1 and ST/M003787/1 and
by an European Research Council Starting Grant {\it ``PDF4BSM''}.

\appendix

\section{PDF correlations}
\label{sec-appendix-correlations}

In this Appendix we illustrate graphically the selection of the region
$\Xi$
Eq.~(\ref{eq:corrinterval}) by the SM-PDF algorithm.

\begin{figure}[t]
\begin{center}
\includegraphics[width=0.45\textwidth]{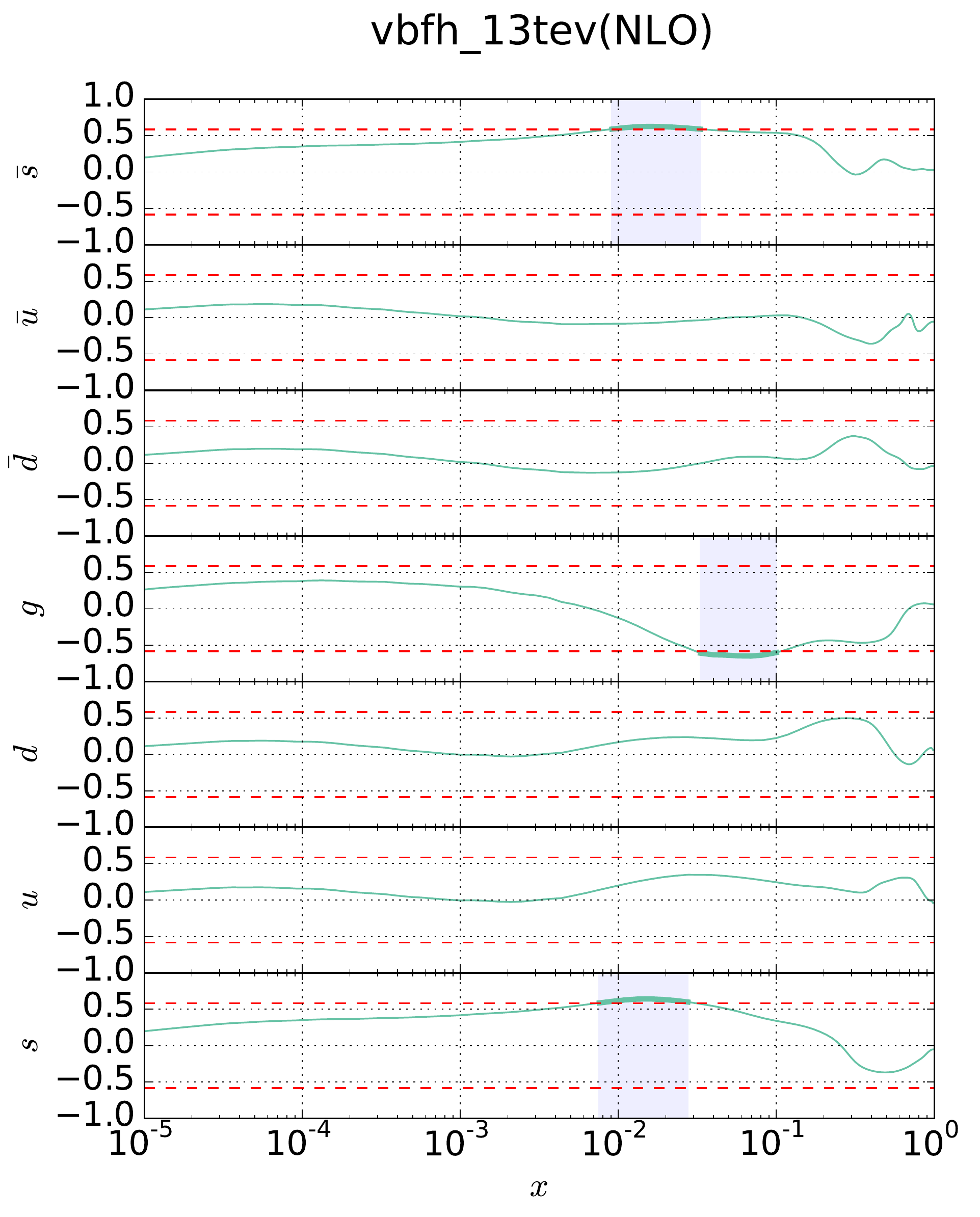}
\includegraphics[width=0.45\textwidth]{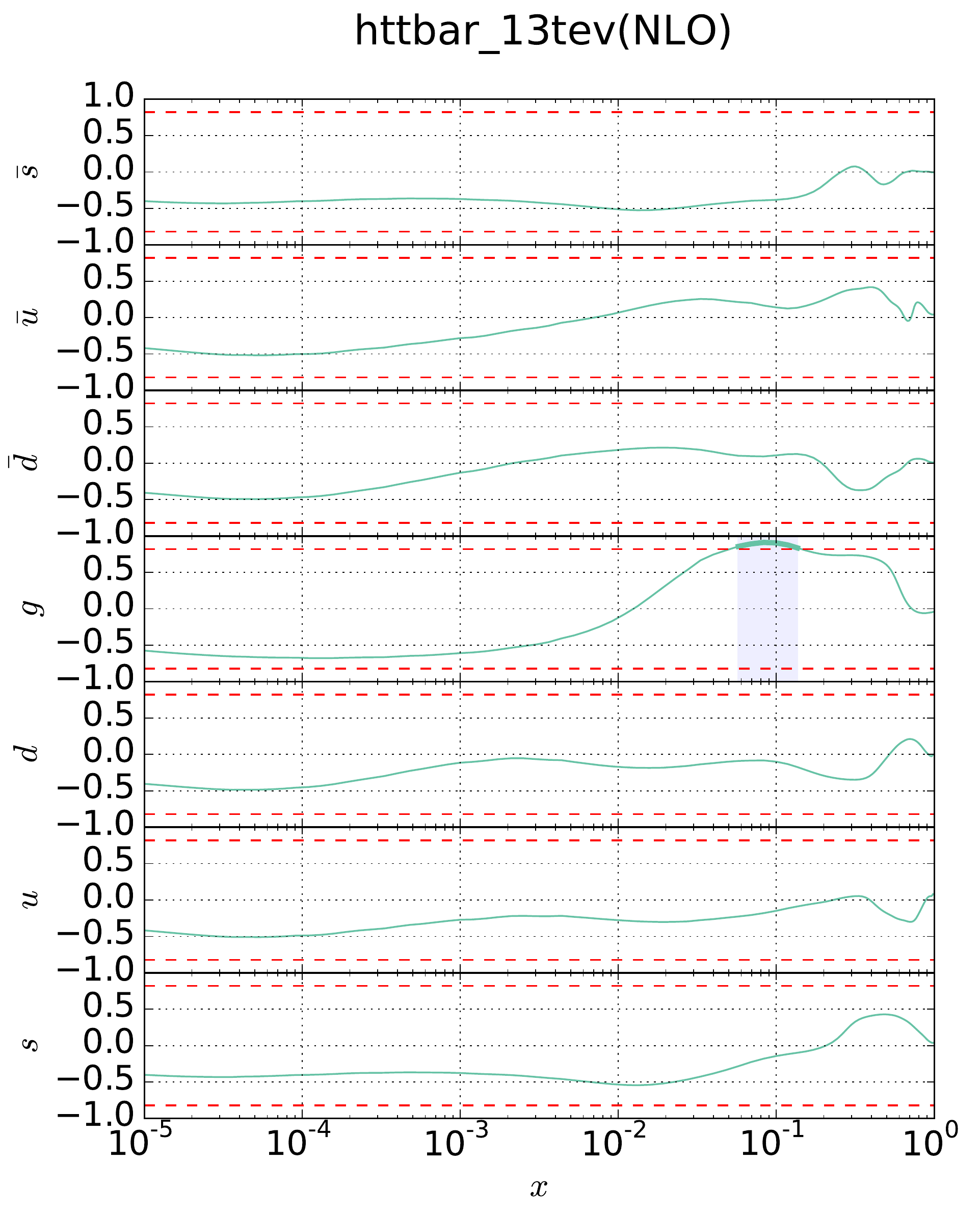}
\end{center}
\vspace{-0.3cm}
\caption{\small \label{fig:pdfcorrelations} Correlation
  Eq.~(\ref{eq:corr_mc})  between  the PDFs and the cross section for
  Higgs production in  vector-boson fusion (left) and associated
with a $t\bar{t}$ pair (right),  as a function of $x$, computed using
  the MC900 NLO PDF set.
The threshold value $\rho = 0.9\rho_{\rm max}$ is shown as a  dashed line, and
  the region in which the correlation coefficient exceeds the threshold,
  $\rho \ge 0.9\rho_{\rm max}$,
  is shown
  as a shaded band.
}
\end{figure}

In Fig.~\ref{fig:pdfcorrelations}
we plot as  a function of $x$ the value of
the correlation Eq.~(\ref{eq:corr_mc}) between
PDFs and the total cross-section 
 for Higgs production in vector-boson fusion (VBF) and in association with
 a $t\bar{t}$ pair, determined   
using MC900 NLO PDFs.
The $\Xi$ region is that in which the correlation exceeds the
 value $\rho = 0.9\rho_{\rm max}$,  shown as a  dashed line in the
 plots, and it is highlighted with a gray band.
The corresponding comparison for Higgs production in gluon fusion
  was shown in Fig.~\ref{fig:corrthreshold}.
  We see  that  $\Xi$ includes
  the gluon around $x\simeq \lp 0.05,0.1\rp$ and the strangeness $s,\bar{s}$ 
  around $x\simeq 10^{-2}$, while for
  $ht\bar{t}$ production it includes the gluon for $x \simeq 0.1$.

\begin{figure}[t]
\begin{center}
\includegraphics[width=0.45\textwidth]{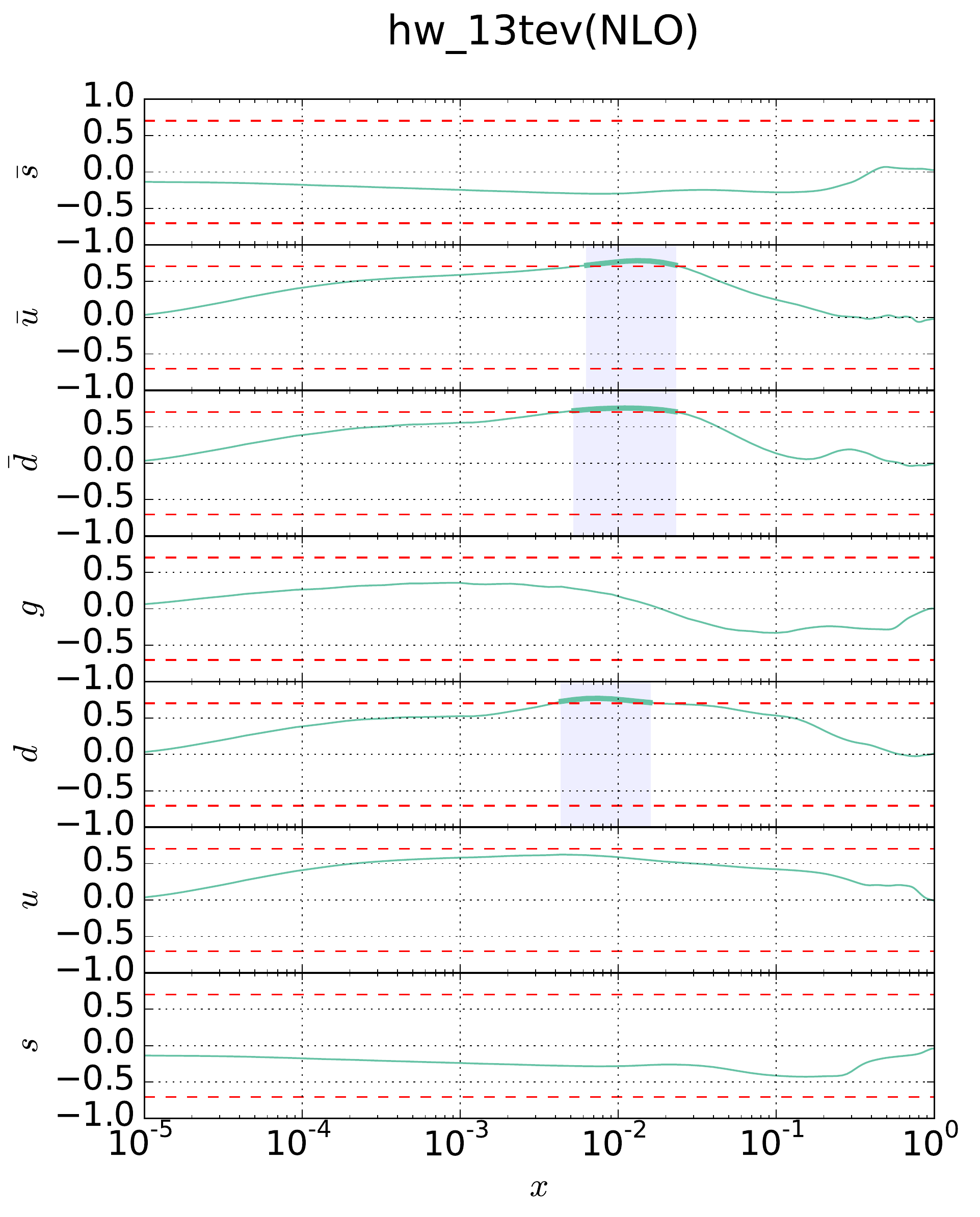}
\includegraphics[width=0.45\textwidth]{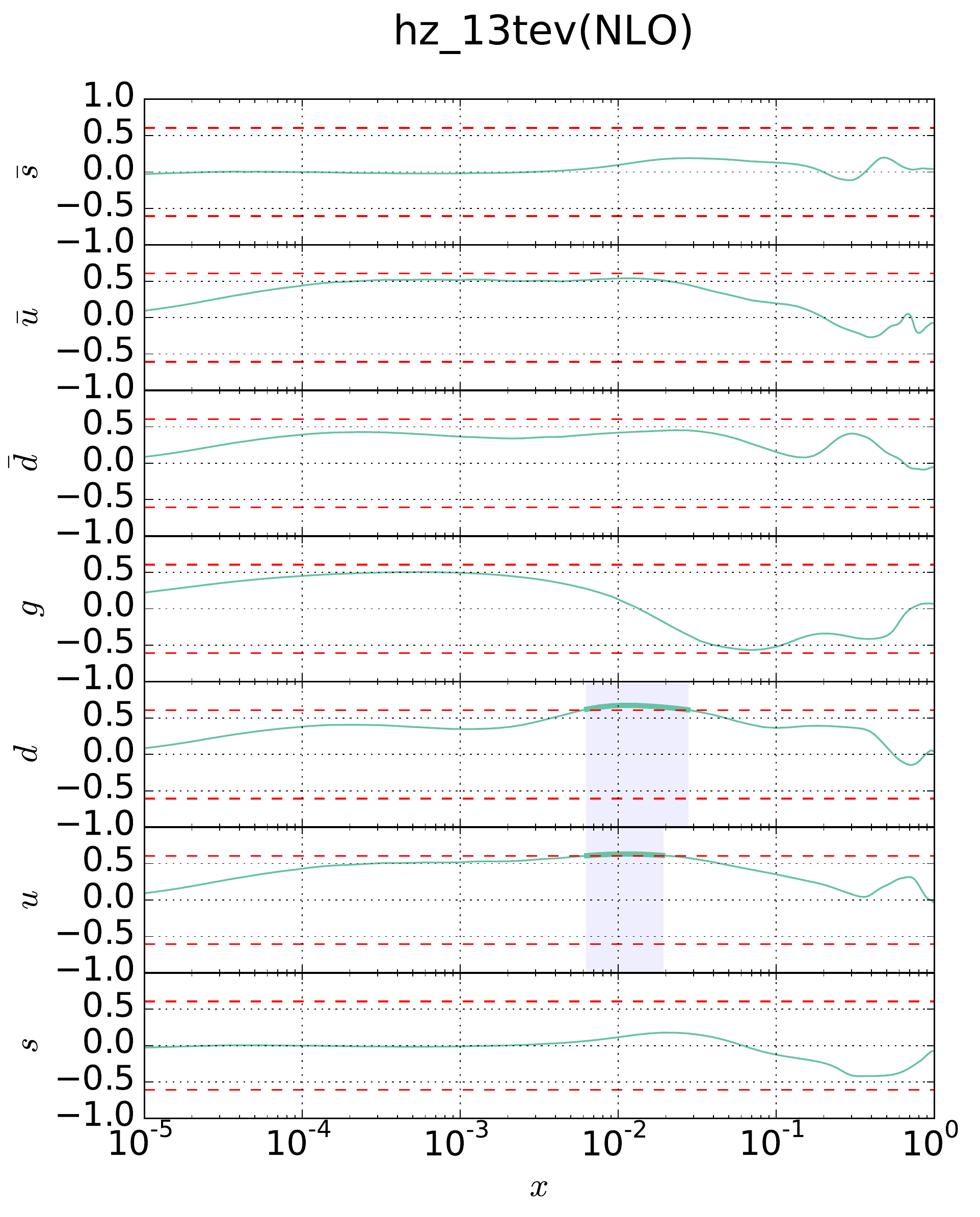}
\end{center}
\vspace{-0.3cm}
\caption{\small \label{fig:pdfcorrelations2}
Same as Fig.~\ref{fig:pdfcorrelations} for associated production
of Higgs bosons with $W$ (left) and $Z$ bosons (right).
}
\end{figure}

The corresponding comparisons for
 Higgs production in association with $W$ and $Z$ bosons
is shown in Fig.~\ref{fig:pdfcorrelations2}.
In this case,  for $hW$ the $\Xi$ region includes
the $\bar{u}$, $\bar{d}$ and $d$ quark PDFs for $x\simeq 10^{-2}$, and for
 $hZ$ production the same region, but for the $u$ and $d$ quark PDFs.

\begin{figure}[t]
\begin{center}
\includegraphics[width=0.45\textwidth]{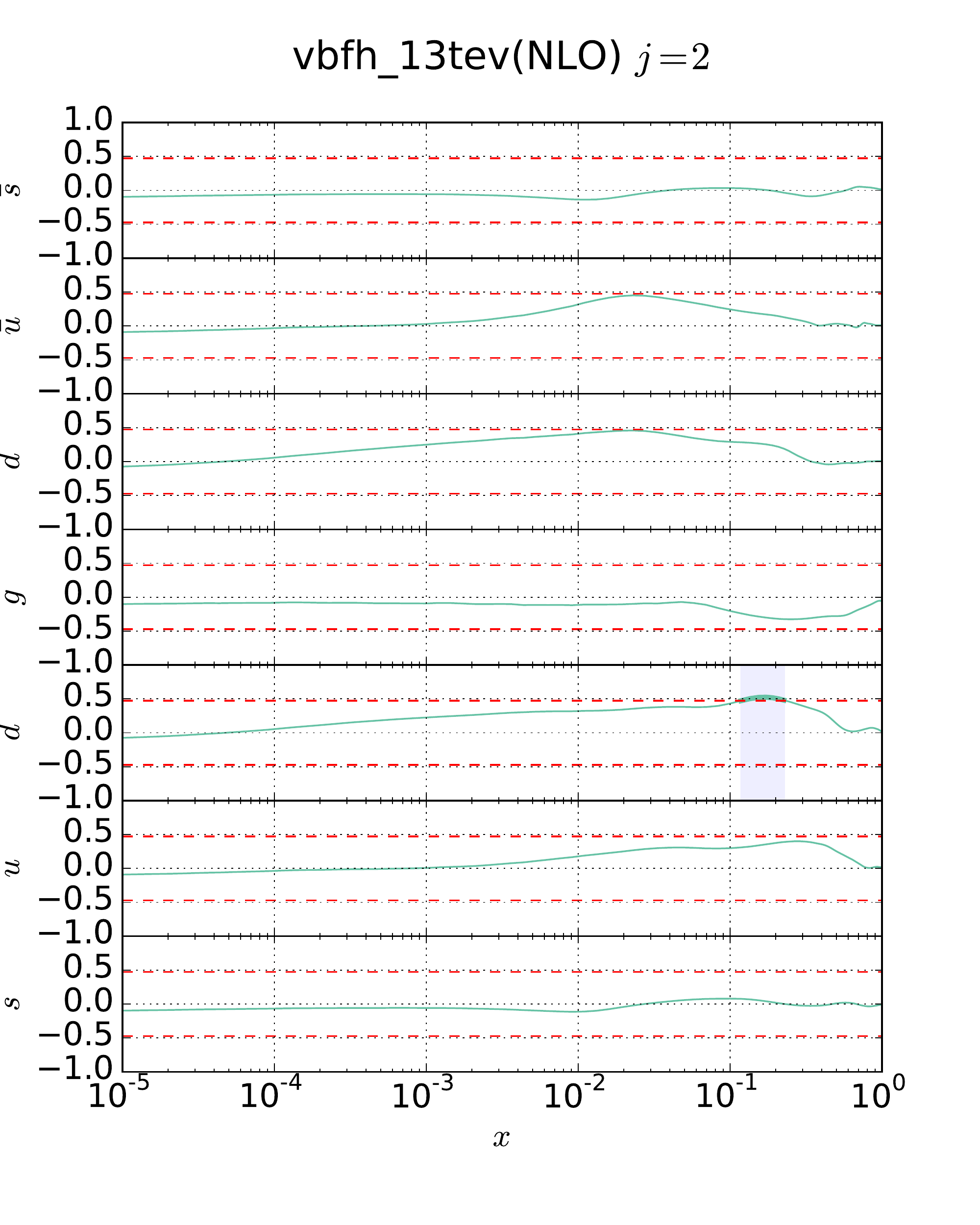}
\includegraphics[width=0.45\textwidth]{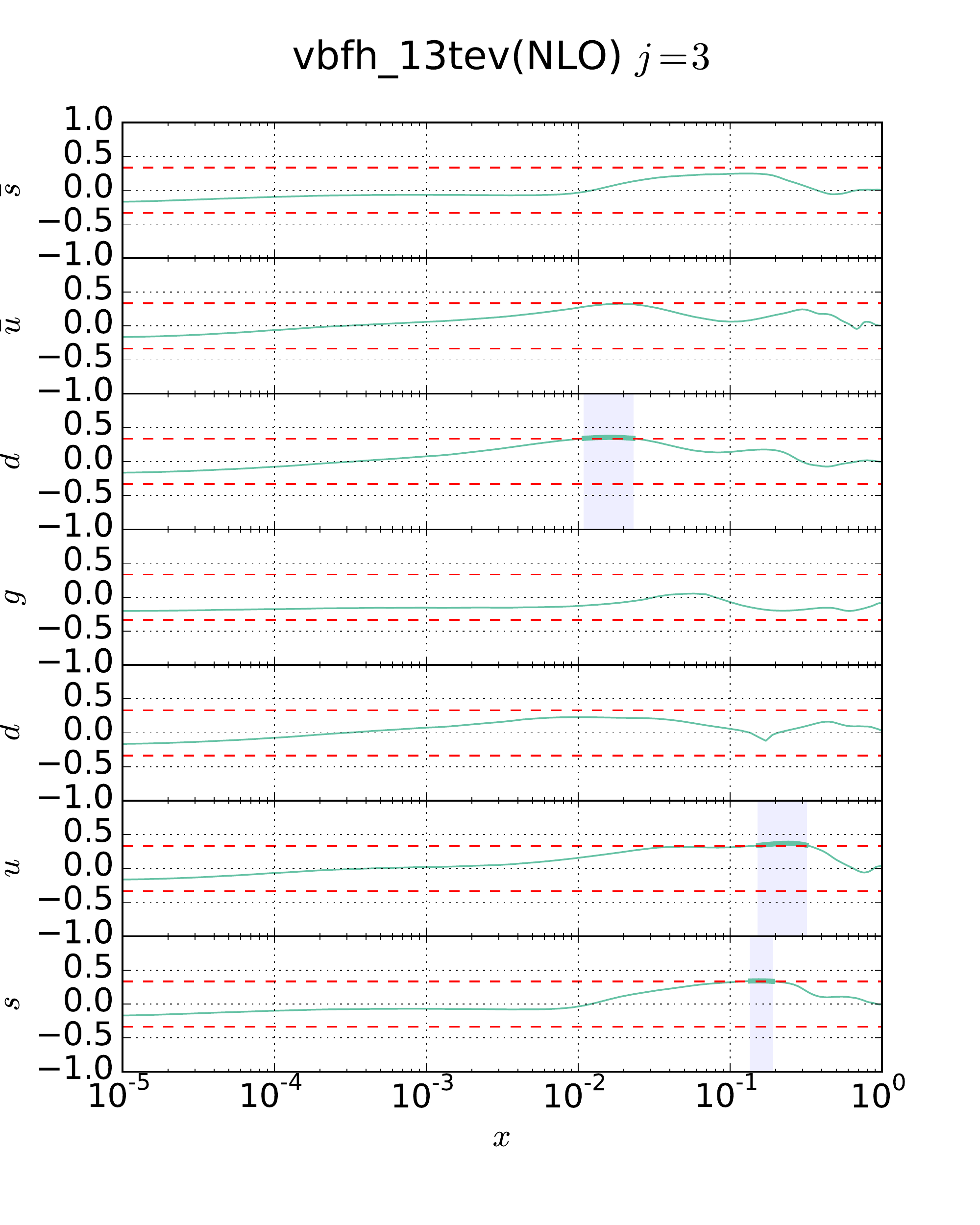}
\end{center}
\vspace{-0.3cm}
\caption{\small \label{fig:corriters2}
Same as the left plot of Fig.~\ref{fig:pdfcorrelations},  but
now at  the second (left) and
third (right) iteration of the SM-PDF algorithm 
}
\end{figure}

\begin{figure}[t]
\begin{center}
\includegraphics[width=0.45\textwidth]{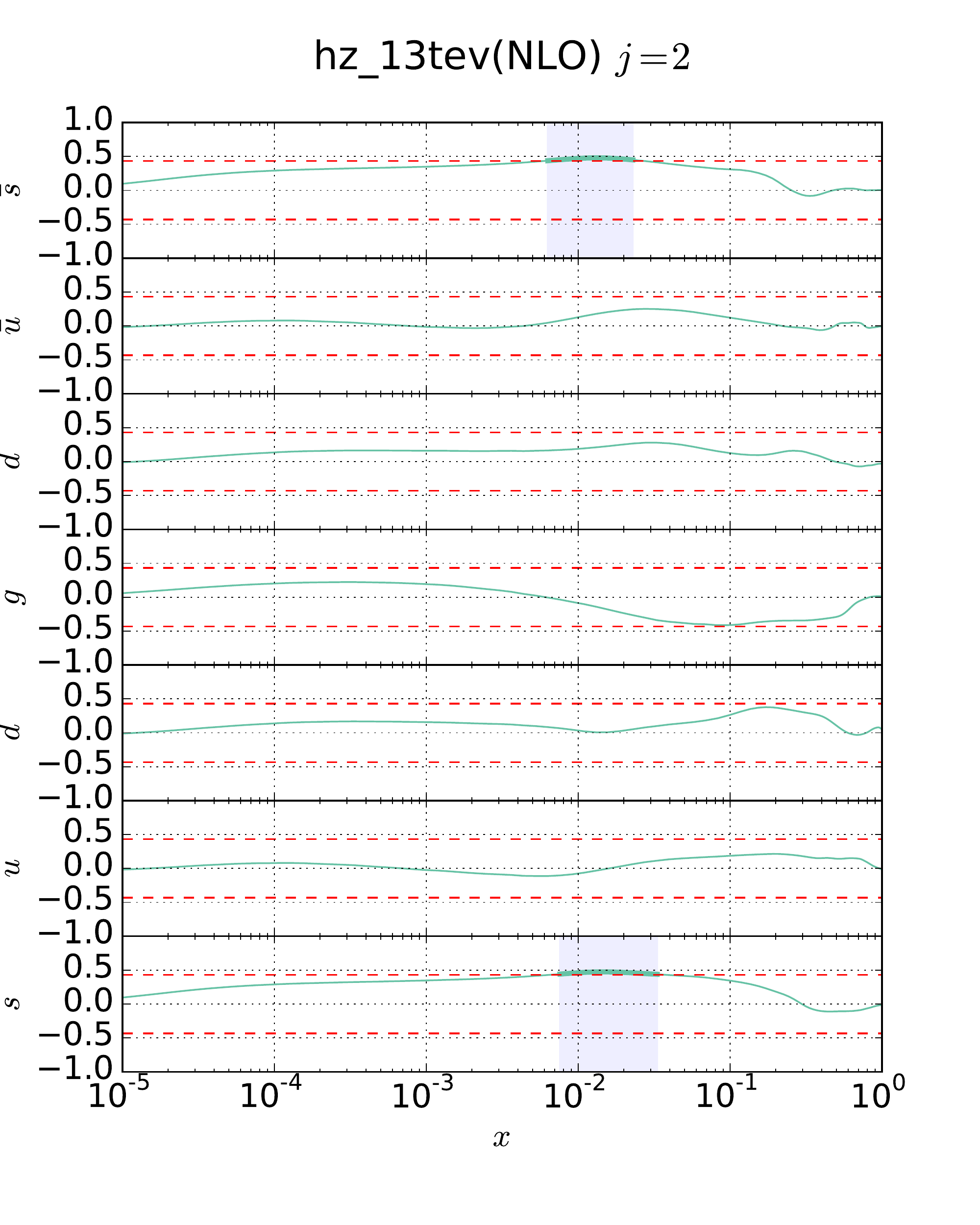}
\includegraphics[width=0.45\textwidth]{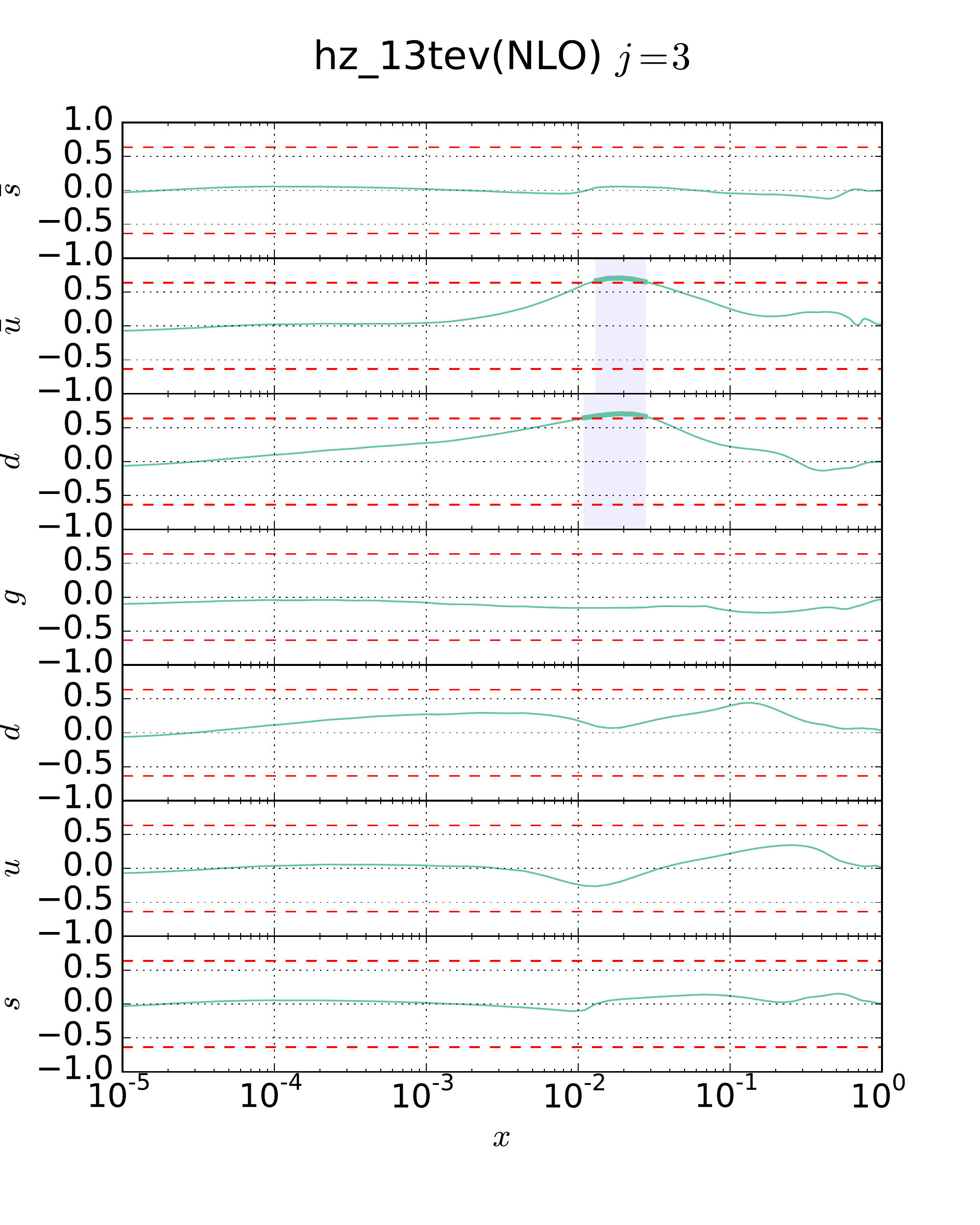}
\end{center}
\vspace{-0.3cm}
\caption{\small \label{fig:corriters} Same as Fig.~\ref{fig:corriters2}.
but now In this case, the results for the first iteration of the algorithm
were shown in
the right plot of
Fig.~\ref{fig:pdfcorrelations2}.
}
\end{figure}

The regions shown in Figs.~\ref{fig:pdfcorrelations}
and~\ref{fig:pdfcorrelations2} are selected at the first iteration of
the SM-PDF algorithm. These are therefore the regions which are needed
in order to determine the most important eigenvector. At the
subsequent iteration, further regions are selected in the orthogonal
subspace. The regions selected at the second and third iterations for
Higgs  production in VBF and $hZ$ production  are respectively shown in
Figs.~\ref{fig:corriters2} (to be compared to the first iteration,
shown in left plot of
Fig.~\ref{fig:pdfcorrelations2})
and in
Fig.\ref{fig:corriters} ((to be compared to the first iteration,
shown in left plot of Fig.~\ref{fig:pdfcorrelations2}).

For VBF  in the second iteration $\Xi$ contains the $d$ PDF at
$x\simeq 0.2$ and the third the $d$ PDF at $x\simeq 0.02$,
and the up and strange PDFs at $x\simeq 0.2$. For 
$hZ$, it contains the strange
PDFs around $x\simeq 10^{-2}$ at the second iteration, and at
 the third iteration the $\bar{u}$ and $\bar{d}$ PDFs for
$x \simeq \lp 0.01,0.05\rp$.
In each case, there is no overlap between regions selected in
subsequent iterations, as it must be because of the projection. The
hierarchy in selection shows which regions and PDFs are increasingly
less important in determining the given cross-section.

\clearpage

\section{Basic usage of the SM-PDF code}
\label{sec-appendix}
While we refer the user to the documentation
bundled in with the SM-PDF code, that will be updated
over time, here we provide an annotated example of a
basic {\tt YAML} configuration file that can be used
to define the inputs to the code.
In particular, the following example of the steering card
is the one used to generate the Higgs SM-PDF set,
constructed using all the processes in Table~\ref{tab:processes_H}
as input.
In addition, the main executable also produces a number of validation
plots such as those presented in
Sect.~\ref{sec:validation}.

Following installation, the SM-PDF code can be executed
using the following
command:
\begin{center}
\tt{smpdf higgs.yaml --use-db}
\end{center}
where the steering card should contain the following information:
\begin{lstlisting}[frame=single, basicstyle=\scriptsize]
# higgs.yaml
# Global parameters that are used unless overwritten by parameters
# inside the action groups
observables: # Indicate the paths to the APPLgrids, and specify the
             # perturbative order in which they have been calculated
   # Higgs
   # Total xsec for Higgs in gluon fusion
   - {name: 'data/higgs/ggh_13tev.root', order: NLO}
   # ggHiggs differential distributions
   - {name: 'data/higgs/ggH_y_13tev.root', order: NLO}
   - {name: 'data/higgs/ggH_pt_13tev.root', order: NLO}
   # Total xsecs for Higgs + W or Z
   - {name: 'data/higgs/hw_13tev.root', order: NLO}
   - {name: 'data/higgs/hz_13tev.root', order: NLO}
   # Total xsecs for Higgs in association with a ttbar pair
   - {name: 'data/higgs/httbar_13tev.root', order: NLO}
pdfsets:
   - MC900_nnlo # LHAPDF6 PDF set to be used as prior in the algorithm
actions:
   - smpdf # Generate the SM-PDF sets from prior PDF set and input observables
   - installgrids # Install the generated sets in the LHAPDF path

#The specification of the actions to actually be performed
#using the above as default
actiongroups:
   - prefix: H05_ #Begin all exported filenames with this prefix
     smpdf_tolerance: 0.05 #Set T to 5% and execute the default
                           #actions above

   - prefix: H10_ 
     smpdf_tolerance: 0.10 #Set T to 10% and execute the default
                           #actions.

   - prefix: compall
     pdfsets: #Change the PDFsets for this actiongroup
        - MCH_nnlo_100
        - H05_smpdf* #Wildcard expansion is supported.
        - H10_smpdf*
        - MC900_nnlo
     actions: #Perform plots and save the data of the convolution.
        - violinplots
        - obscorrplots
        - ciplots
        - savedata
     base_pdf: MC900_nnlo #Plot values relative to this PDF.

\end{lstlisting}
No additional settings need to be modified.
By default, the code will also output the generated SM-PDF set
directly in the {\tt LHAPDF6} format.

\providecommand{\href}[2]{#2}\begingroup\raggedright\endgroup


\end{document}